# DNA methylation variation in Arabidopsis has a genetic basis and shows evidence of local adaptation

Manu J. Dubin[1*], Pei Zhang[1,2*], Dazhe Meng[1,2*], Marie-Stanislas Remigereau[2*], Edward J. Osborne[3], Francesco Paolo Casale[4], Phillip Drewe[5,6], André Kahles[5,6], Bjarni Vilhjálmsson[1], Joanna Jagoda[1], Selen Irez[1],Viktor Voronin[1], Qiang Song[2], Quan Long[1], Gunnar Rätsch[5,6], Oliver Stegle[4], Richard M. Clark[3,7], Magnus Nordborg[1,2]

1. Gregor Mendel Institute, Austrian Academy of Sciences, Vienna, Austria.
2. Molecular and Computational Biology, University of Southern California, Los Angeles, California, USA.
3. Department of Biology, University of Utah, Salt Lake City, UT, USA.
4. European Molecular Biology Laboratory, European Bioinformatics Institute, Wellcome Trust Genome Campus, Hinxton, Cambridge, United Kingdom.
5. Friedrich Miescher Laboratory, Max Planck Society, Spemannstraße 39, 72076 Tübingen, Germany.
6. Memorial Sloan-Kettering Cancer Center, New York, NY, USA.
7. Center for Cell and Genome Science, University of Utah, Salt Lake City, UT, USA.

## Abstract:

Epigenome modulation in response to the environment potentially provides a mechanism for organisms to adapt, both within and between generations. However, neither the extent to which this occurs, nor the molecular mechanisms involved are known. Here we investigate DNA methylation variation in Swedish *Arabidopsis thaliana* accessions grown at two different temperatures. Environmental effects on DNA methylation were limited to transposons, where CHH methylation was found to increase with temperature. Genome-wide association mapping revealed that the extensive CHH methylation variation was strongly associated with genetic variants in both cis and trans, including a major trans-association close to the DNA methyltransferase CMT2. Unlike CHH methylation, CpG gene body methylation (GBM) on the coding region of genes was not affected by growth temperature, but was instead strongly correlated with the latitude of origin. Accessions from colder regions had higher levels of GBM for a significant fraction of the genome, and this was correlated with elevated transcription levels for the genes affected. Genome-wide association mapping revealed that this effect was largely due to trans-acting loci, a significant fraction of which showed evidence of local adaptation.

## Impact statement:

These findings constitute the first direct link between DNA methylation and adaptation to the environment, and provide a basis for further dissecting how environmentally driven and genetically determined epigenetic variation interact and influence organismal fitness.

## Main

To better understand how genotype and environment interact to affect DNA methylation and transcription, we grew 150 *A. thaliana* accessions from Sweden (Long et al., 2013) in two different environments, 10 C and 16 C, chosen because they lead to very different flowering behavior (Atwell et al., 2010). Relying on existing genome sequence information (Long et al., 2013), methylome- and transcriptome-sequencing data were generated (see methods).

In plants, DNA methylation occurs on cytosines in the CG, CHG, and CHH contexts (where H is any nucleotide except for C), each of which is catalyzed by independent pathways (Finnegan et al., 1998; Stroud et al.). Consistent with previous results (Eichten et al., 2013; Li et al., 2014; Schmitz et al. 2013; Vaughn et al., 2007) we found considerable variation between accessions regardless of context, even at the level of genome-wide averages (Fig 1a). Temperature, on the other hand, did not appear to affect genome-wide CG or CHG methylation, but had a significant effect on CHH methylation, levels of which were 14% higher at 16 C than at 10 C, on average (Fig. 1a). To investigate the genetic basis of DNA methylation, we performed genome-wide association studies (GWAS) using different facets of average methylation as the phenotype. For global CG and CHG methylation, no associations reached genome-wide significance, while for CHH methylation a clear peak of association was observed on chromosome 4 (Supplementary Fig S 1. a-c). The association was even more significant when restricting attention to average CHH methylation of large transposons (Fig. 1b), in agreement with the notion that this type of methylation mostly occurs in transposons in plants (Finnegan et al., 1998).

The association centered around a SNP at 10,459,127 on chromosome 4, 38 kb downstream from the locus AT4G19020, which encodes a homolog of the CHG methyltransferase chromo-methylase-3 (Lindroth et al., 2001) that has recently been shown to catalyze both CHH and CHG methylation on transposons, and is thus an excellent candidate (Stroud et al., 2014; Zemach et al., 2013). A multi-locus mixed model (Segura et al., 2012) that included the identified SNP (*CMT2a*) as a fixed effect revealed another SNP downstream of *CMT2*, at position 10,454,628 (*CMT2b*), 8 kb closer to *CMT2* than *CMT2a*, and in complete linkage disequilibrium with it (i.e., the non-reference alleles at *CMT2a* and *CMT2b* are never seen together). Repeating the GWAS with both *CMT2a* and *CMT2b* as cofactors identified no further loci (Supplementary Fig. S1d-f). Both non-reference alleles are common in southern Sweden, but are also found in the north (22.6% vs 9.5% and 30.6% vs 7.9% for *CMT2a* and *CMT2b*, respectively). Accessions with the non-reference *CMT2a* allele have on average more CHH methylation on transposons than those with the reference haplotype (p = 1.1e-03), while those with the non-reference *CMT2b* allele have lower levels of CHH methylation than the reference haplotype (p = 8.1e-03; Fig. 1c). The associations were readily confirmed using an F2 population generated by crossing accessions with the *CMT2a* and *CMT2b* non-reference alleles (Fig. 1d-e). No significant differences in CMT2 mRNA levels were observed between the alleles in our data and limited Sanger sequencing of cDNA showed no evidence of splicing variants (although, as will be discussed below, we did detect a putative rare null allele). Several non-synonymous polymorphisms in the methyltransferase and BAH domains of CMT2 were detected (Supplementary File1) but they do not explain the phenotype as well as the *CMT2a* and *CMT2b* SNPs.

The effect of genetic variation on local CHH methylation was examined by calculating the methylation level in 200 bp windows across the genome, and running GWAS for the 200,000 differentially methylated regions (DMRs; see Methods) that varied most between individuals. 18,057 DMRs had at least one statistically significant association (p-value < 0.05 after Bonferroni correction). After merging DMRs within 100 bp of each other, 10,812 distinct DMRs remained. This is approximately 10-fold more than previously reported for a worldwide sample(Schmitz et al., 2013), however, reanalysis of these data using our approach suggests that the difference is largely due to methodology (not shown). 45% of the DMRs had a significant *cis*-association within 100 kb, while the rest showed evidence of *trans*-regulation, including the dramatic effect of *CMT2* on chromosome 4 which accounted for approximately 20% of all significant associations (Fig. 2a).

The effect of temperature on CHH methylation could also be seen at the local level. We defined "temperature DMRs" by looking for windows that differed significantly between temperatures. Comparing 16 C to 10 C, each accession on average gained CHH methylation at ~400 temperature DMRs and lost it at ~200 temperature DMRs (false discovery rate= 0.05). CHH methylation is associated with transposable elements (TEs; Finnegan et al., 1998), and in agreement with this, 79% of the temperature DMRs where methylation was gained at 16 C were located within 500 bp of an annotated TE (with 60% directly overlapping one). These temperature DMRs were enriched in small subset of TEs (835, or 2.7% of total, permutation based p-value =0.05) that were more highly methylated than other transposons, but with lower methylation levels immediately adjacent (Fig. 2b). Compared to TEs without temperature DMRs, these "variable" TEs also tended to be euchromatic (Supplementary Fig. S2a), highly expressed (Supplementary Fig. S2b), and recently inserted ("evolutionarily young" TE insertions for which orthologs are not present in in *A. lyrata* (Zhong et al., 2012) comprised 75% of the variable TEs vs 68% of non-variable TEs). At the super-family level, members of the *SINE*, *SINE*-like, *Helitrons* and *Mutator*-like DNA TE superfamilies were over-represented among the variable transposons, and at the family level, 36 families were over-represented, including the *AtREP*, *Vandal* and *HAT* DNA transposons, as well as *COPIA78/ONSEN* and *META1* retroelements (Supplementary Table S1). Interestingly, *COPIA78* has been shown to become active in response to heat stress (Ito et al., 2011; Pecinka et al., 2010) apparently due to heat-shock promoter elements in its LTR regions (Cavrak et al. 2014).

In order to gain further insight into the mechanisms responsible for variation in CHH methylation, we bisulfite-sequenced knockout lines of CMT2 (SAIL_906_G03) and DCL3 (*dcl3-5* (Daxinger et al., 2009), a component of the RdDM pathway), and identified 10,138 DCL3-dependant DMRs and 33,422 CMT2-dependent DMRs (Stroud et al., 2013). As expected under the assumption that CMT2 is responsible for the massive GWAS peak on chromosome 4, this peak remains for CMT2-dependent DMRs, but not for DCL3-dependent DMRs (Supplementary Fig. S1c-d). Furthermore, while CHH methylation varied with temperature at both DCL3- and CMT2-dependent DMRs (Fig 2c), 33% of the former overlapped with the previously identified temperature DMR's, compared to only 3% of the latter, suggesting that much of the temperature variation in CHH methylation is due to components of the RdDM pathway. This result is consistent with previous findings showing that RNA silencing is less active at lower temperatures (Romon et al., 2013).

Interestingly, we observed one accession from northern Sweden, TAA-03, with almost undetectable levels of CHH methylation at CMT2-dependant DMRs (Fig 2c). Further investigation suggested that it has a deletion or rearrangement in CMT2, as we were unable to map reads between positions 2813 and 4944 (intron 7 to exon 16, Supplementary Fig. S2e). Sanger-sequencing indicates the insertion of a stretch of TC dinucleotide repeats of at least 330bp. The same deletion appears to be present in three more accessions from northern Sweden (TAA-14, TAA-18, and Gro-3). This adds to the emerging picture that CMT2 harbors considerable functional variation, including natural null alleles, and is reminiscent of what was previously observed for its homolog CMT1 (Henikoff and Comai, 1998).

To quantify the regulation of DMRs, we partitioned the variance of CHH methylation into environmental (E), *CMT2*, *CMT2* X E, *cis*, *cis* X E, *trans*, and *trans* X E using a mixed model (Fig. 2d). The analysis confirmed substantial *cis* and *trans* effects, with the environment modulating the genetic effects rather than having a major direct effect. A possible explanation is that SNPs tag polymorphic TE insertions, with the insertion allele being silenced in a temperature-sensitive manner.

It has recently been suggested that natural variation in CMT2 is associated with adaptation to climate (Shen et al., 2013). Given the sensitivity of CHH methylation to growth temperature observed here, we next investigated the correlation between DNA methylation and the climate of origin (Hancock et al., 2011). While CHH methylation was moderately correlated with photosynthetically active radiation (PAR) in spring (Pearson's r = 0.38), and CHG showed correlation with aridity (r = 0.35) and the number of frost-free days (Pearson's r = 0.30), by far the strongest signal was a strong positive correlation between CG methylation and latitude (Pearson's r = 0.70), as well as with a number of environmental variables that co-vary with latitude in our sample, such as minimum temperature and daytime length (Supplementary Table S2, Fig. 3a). As a result of the strong latitudinal correlation, accessions originating from northern Sweden (minimum temperature below -10 C) had on average 11% higher global CG methylation compared to those from the south (Fig. 3a). The correlation appears to be driven by gene body methylation (GBM): as the correlation for CG methylation on transposons was much weaker (Supplementary Fig. S3a,b). Because the methylation variation observed for genes with average CG methylation below 5% appeared mostly to be noise (see Methods), we classified genes into "unmethylated" and "having GBM" using this as a cutoff. We also eliminated genes showing a transposon-like pattern of methylation in which not only CG, but also the CHH and CHG contexts are highly methylated (Zemach et al., 2013). In what follows, we use GBM to refer only to gene body CG methylation for this filtered set.

GBM primarily occurs on long, evolutionary conserved genes that tend to be moderately-to-highly expressed, and is positively correlated with gene expression (Takuno and Gaut, 2012; Zilberman et al., 2007). Genes with higher GBM tended to be more highly expressed in our data as well, and — more interestingly — accessions with higher average GBM showed higher average expression of methylated genes (Supplementary Fig. S3c). Given that northern accessions had higher GBM, this meant that genes with GBM were on average more highly expressed in northern than in southern accessions, while unmethylated genes showed little difference (Fig. 3b).

As for CHH DMRs, the genetic basis of GBM was examined using a variance-component approach (Fig. 4a). The results were dramatically different: relative to CHH methylation, the *trans* effects for GBM are massive, and the environment appears to have no effect (in agreement with the observation that only CHH methylation levels vary with temperature, see Fig. 1a). To identify the genes responsible, we also performed GWAS for each gene with GBM (Fig. 4b). A total of 3241 significant associations were found for 2315 genes. 43% of these genes had a significant *cis*-association (within 100 kb of the gene of interest), which could represent local variants affecting methylation directly, or indirectly by affecting gene expression (Gutierrez-Arcelus et al., 2013). No evidence for major trans-acting loci like *CMT2* was found, but 69% of all significant associations were in *trans*. A comparison of the direction of the effect of GBM-associated SNPs in *cis* and *trans* revealed a striking pattern (Fig. 4c). While the non-reference alleles of *cis*-SNPs were 1.18 times more likely to be associated with decreased rather than increased GBM ($p$ = 2.01e-04), the non-reference alleles of *trans*-SNPs were 3.48 times more likely to be associated with increased GBM ($p$ = 2.2e-16), and the non-reference alleles at the 15 *trans*-SNPs that were associated with GBM at five or more genes were always positively correlated (Fig. 4c). Furthermore, while *cis*-SNPs showed a wide distribution of allele frequencies similar to random SNPs, *trans*-SNPs showed a much more limited distribution of frequencies (Supplementary Fig. S4a) and were also much more strongly correlated with latitude (Fig. 4d). The correlation between GBM and latitude thus appears mostly to be due to *trans*-acting SNPs.

The 15 highly associated *trans*-SNPs were largely limited to northern Sweden, and were in strong linkage disequilibrium with each other (Supplementary Fig. S4b-c). *A. thaliana* from northern Sweden show signs of multiple strong selective sweeps (Long et al., 2013) and harbors many polymorphisms that appear to be involved in local adaptation (specifically to minimum temperature; Hancock et al., 2011). The 15 SNPs were more than 9-fold over-represented in the previously identified sweep regions (empirical p-value = 5.1e-03) and over 5-fold over-represented within 2 kb of SNPs in the 1% tail of those associated with minimum temperature (Hancock et al., 2011) (empirical p-value = 3.1e-04), Supplementary Table S3). The ancestral state could be determined for 10 of the 15 SNPs, and in 8 of these cases, the non-reference allele was derived, further supporting sweeps in northern Sweden.

Identifying the causal variants is challenging, however, because the peaks of association are broad (Fig. 4b), as expected for polymorphisms involved in local adaptation (Atwell et al., 2010; Zhao et al., 2007). A gene-ontology analysis of genes within 100 kb (the average size of the sweep regions (Long et al., 2013)), of the 15 *trans*-SNPs found enrichment of loci associated with mRNA transcription (GO0009299, p-value = 2.62e-03). Genes involved in epigenetic processes are not captured well by standard gene-ontology, but we found that genes from the plant chromatin database (www.chromdb.org/) were significantly overrepresented in these regions as well (permutation p-value = 0.012; Supplementary Table S4).

We also looked for genes whose expression variation is consistent with a causal role. We identified 68 genes within 100 kb of one of the 15-trans SNPs whose expression is highly correlated (Wilcoxon test $p < 0.001$) with the adjacent SNP after correction for population structure (Supplementary Table S5). No significant enrichment of Gene Ontology terms was observed among these genes, and manual inspection identified no proteins directly involved in DNA methylation. Instead, a number of proteins involved in the regulation of gene expression and/or chromatin accessibility were present (Supplementary Table S5). This may suggest that the increased gene-body methylation observed in the north is not directly due to increased DNA methylation, but may be caused by increases in gene expression driven either by differences in transcription factors networks or chromatin compaction. Identification of the causal variants behind this phenomenon should provide insight into how plants adapt to their local environment.

In conclusion, genes with GBM are generally up-regulated and more heavily methylated in northern accessions, and the change appears to be due to *trans*-acting polymorphisms that have been subject to directional selection. The candidate regions show an overrepresentation of genes involved in transcriptional processes. We also found that CHH methylation of TEs is temperature sensitive, and identified a major *trans*-acting controller, *CMT2*. Taken together, these observations suggest that local adaptation in *A. thaliana* involves genome-wide changes in fundamental mechanisms of gene regulation, perhaps as a form of temperature compensation.

# Materials and Methods

## 1 Raw data generation

### 1.1 Plant growth

A diverse set of 150 Swedish accessions were sown on soil and stratified for 3 days at 4 C in the dark. They were then transferred to environmentally controlled growth chambers set at 10C or 16C under long day conditions (04:00-20:00) and individual seedlings were transplanted to single pots after one week. When plants attained the 9-true -leaf stage of development, whole rosettes were collected between 15:00 and 16:00 hours and flash frozen in liquid nitrogen.

In addition, a cross between the T550 and Brösarp-11-135 accessions was created and F2 seed obtained. 36 individual F2 lines were grown in the same manner as the accessions.

### 1.2 RNA-seq library preparation

For each accession, 3 plants were pooled and total RNA was extracted by TRIzol (Invitrogen 15596-018), DNase treated and mRNA purified with oligo dT Dynabeads (Life Technology). RNA was then fragmented using Ambion Fragmentation buffer and first and second strand cDNA synthesis was carried out using Invitrogen kit 18064-071. The ends of sheared fragments were repaired using Epicentre kit ER81050. After A-tailing using exo-Klenow fragment (New England Biolabs, MA, NEB M0212L), barcoded adaptors were ligated with Epicentre Fast-Link DNA Ligation Kit (Epicentre LK6201H). Adaptor-ligated DNA was resolved on 1.5% low melt agarose gels for 1 hour at 100V. DNA in the range of 200 -250 bp was excised from the gel and purified with the Zymoclean Gel DNA recovery kit (Zymo Research). The libraries were amplified by PCR for 15 cycles with Illumina PCR primers 1.1 and 1.2 with Phusion polymerase (NEB F-530L).

Single-end 32 bp sequencing was performed at the University of Southern California

Epigenome Center on an Illumina GAIIx instrument using 4-fold multiplexing.

### 1.3 BS-seq library preparation

For each accession two individual plants were pooled and total DNA was extracted using CTAB and phenol-chloroform. Approximately two micrograms of genomic DNA was used for BS-seq library construction and sequencing (92 bp paired-end) by BGI.

## 2 Sequence analysis

### 2.1 Genome sequences

Illumina sequencing data from 180 published Swedish genomes (Long et al., 2013) were combined with sequencing data from another 79 (1001genomes.org), which had been processed using the same pipeline to yield polymorphism data for a total of 259 accessions (including those used for BS-seq and RNA-seq here). Linkage disequilibrium calculated using the R package LDHeatmap (version 0.9.1; Shin et al., 2006).

### 2.2 RNA-seq data processing

#### 2.2.1 Read mapping

After demultiplexing, 36bp RNA-Seq reads were trimmed from barcodes (4nt) and mapped to the TAIR10 reference genome including known variation with the PALMapper aligner (Jean et al., 2010) using a variant-aware alignment strategy. Two different sources of variants were used: 1) single nucleotide variants (SNV) and structural variants (SV) from genome sequencing (2.1) and 2) SNVs and SVs called in an initial alignment round of the RNA-Seq reads to the TAIR10 reference genome with PALMapper (relevant parameters: -M 4 -G 4 -E 6 -I 25000 -NI 1 -S). For both sources of variants we applied stringent filter criteria to reduce false calls: 1) genome variants had to appear in at least 40 strains with a minor allele count of at least 5 strains, 2) RNA-Seq variants had to be confirmed by at least 2 alignments within the same strain and had to have less than factor 2 many non-confirming alignments within the same strain. Variants from both sources were integrated into one file that was used for a second, variant-aware alignment round with PALMapper (relevant parameters: -M 2 -G 0 -E 2 -I 5000 -NI 0 -S). In variant-aware alignment mode, PALMapper builds an implicit representation of the reference genome that reflects all possible variant combinations that exist for a genomic region. The output is automatically projected to the TAIR10 coordinate system. To account for reads ambiguously mapping to multiple locations in the genome, we used a custom python script to remove all reads that showed at least one mapping additional to the best hit with the same edit distance. Additional hits were only counted as ambiguous, if they differed at least 3nt in start and stop coordinates to the best hit. On average, 5.7M reads were mapped per sample after removal of ambiguous reads. Low complexity libraries with less than 30% of mappable reads or samples with less than 800,000 mappable reads (6 in total) where excluded from further analysis.

### 2.2.2 Sample validation

To correct for possible sample or data mix-ups, SNP were called from the RNA-seq alignments using a custom python script and compared to independently collected SNPs from the Arabidopsis 250K SNP array (Kim et al., 2007). Samples not matching the expected genotype were reassigned to the correct genotype where possible or otherwise excluded from further analysis.

### 2.2.3 Filter for gene expression quantification

We quantified gene expression by counting the number of reads that were longer than 24 bp and that mapped to genes on all non-chloroplast and non-mitochondrial chromosomes. To obtain a stable quantification, we only used those reads which were uniquely mapped into the exonic regions of genes. Furthermore, we required that the reads did not map completely into regions where two genes overlap in order to avoid mixing quantifications of different genes. In the later text we will refer to this estimate as the raw expression estimate.

We also quantified the gene expression when additionally accounting for structural variants, alternative splicing and repetitive sequences that can all bias gene expression quantification. This estimate will be referred to as sv-corrected expression. For this quantification we additionally filtered for reads that start in an insertion or deletions and their two neighbouring bases, that mapped into regions that are not contained in all transcripts of a gene and reads which were mapped completely into regions which are repetitive based on a 50 bp window.

#### 2.2.4 Quantification per ecotype and environment.

After filtering (see 2.2.3), there were 499 RNA-Seq libraries left for analysis. Next, we merged libraries per ecotype and environment, yielding 323 unique merged RNA-Seq libraries for a unique ecotype and environment (160 in 10C, 163 in 16C).

#### 2.2.5 Estimation of library size and abundance estimates

We followed the low level normalization proposed by (Anders and Huber, 2010), jointly applied to the set of expression estimates across ecotypes and environmental backgrounds. First, we estimated effective library sizes as the median expression estimates across all genes. Based on this, we derived correction factors to adjust individual libraries for differences in size.

#### 2.2.6 RKPM values

Library-size adjusted raw counts were used to obtain standard read counts per million expression estimates for each gene.

### 2.3 BS-seq data processing

#### 2.3.1 Read mapping

Reads were aligned as previously described (Dinh et al. 2012) to a modified pseudo-reference chromosome in which SNPs were inserted into the TAIR10 reference genome using NextGenMap (version 0.4.3; Sedlazeck et al. 2013) allowing up to 10% mismatch between the reads (-i 0.90) and the reference sequence and discarding reads that map equally well to more than one genomic location or have less than 45 nucleotides mapping without error to the reference sequence (-R 45). Average coverage was 12.6 X.

To correct for sample or data mix-ups, the raw data was also aligned to the first chromosome of the Columbia-0 TAIR reference genome as described above and SNP calling performed using the BISsnp package (Liu et al. 2012). The polymorphism data were then compared to data from genome sequencing (1001genomes.org). Accessions that did not have the highest similarity to the expected genotype were excluded from further analysis.

### 2.3.2 DNA methylation analysis

Methylation was estimated individually for each cytosine using a python script provided with the BSMAP software package (Xi and Li, 2009). Conversion efficiency was estimated from the fraction of methylated cytosines in chloroplasts using the R software package (www.r-project.org, version 2.15.2). After eliminating one outlier, the samples had conversion efficiencies ranging from 99.25%–99.80% (mean=99.59%). Genome wide average methylation levels were calculated separately for the CG, CHG and CHH contexts. The average variance between 11 biological replicates was 2.2%, 3.2% and 7.3% for CG, CHG and CHH methylation respectively, while for identical genotypes grown at different temperatures (111 pairs) CG, CHG and CHH methylation variance was 2.7%, 4,6% and 15.9% respectively. The variance in genome wide methylation levels for the 152 accessions grown at 10 C was respectively 7.6%, 9.2% and 13.2% for CG, CHG and CHH methylation, while for the 121 accessions grown at 16 C genome wide CG, CHG and CHH methylation varied 8.5%, 9.5% and 14.3% respectively.

The Bioconductor package Repitools (version 0.6.0; Statham et al., 2010) was used to average methylation over genomic features of interest (e.g., all genes, all transposons over 4 kB or a subset of transposons of interest). Pairwise DMRs were called individually for each accession using the R software package methylKit (version 0.5.6; Akalin et al., 2012) using a window size of 200 bp, an FDR rate of 0.05 and a minimum fold change of 0.3. Overlap of DMRs with (TAIR10) genomic features such as transposons and genes was calculated using the Bioconductor package ChIPpeakAnno (version 2.8.0; Zhu et al., 2010).

For each accession, methylation data was smoothed independently for each context using the Bioconductor package BSmooth (version 0.4.5; Hansen et al., 2012) using the default settings. Average methylation was then calculated for 200bp windows across the genome.

# 3 Population genetic analysis

## 3.1 GWAS

Linear mixed models that correct for confounding by the genetic background using a kinship matrix calculated from genetic data were used throughout (Kang et al., 2010; Segura et al., 2012). To examine the effect of genotype on local CHH methylation variation, DMRs were defined by filtering the 200 bp methylation windows to remove those containing missing data (no coverage) in one or more accessions, then selecting the $10^5$ remaining windows with the greatest variance in DNA methylation. For GBM, genes were filtered to remove those that had more than 0.05 average CHG methylation or less than 0.05 average CG methylation across the accessions (Supplementary Fig. 5)

## 3.2 Variance component analysis

To investigate the relative contributions to methylation differences of *cis* and *trans* genetic loci as well as the environment, we used the package LIMIX (https://github.com/PMBio/limix), which efficiently estimates variance components using linear mixed models. The *cis*-window considered here consists of 50 kb on either side of the DMR, whereas *trans* is the rest of the genome.

For each DMR, we considered a linear mixed model with a fixed effect for the environment (10C and 16C), and random effects for *CMT2*, *cis* and *trans* genetic contributions. Let us denote the number of samples with N, the number of environments with E (E = 2), and the number of genomic features with S. Let us focus on a single DMR and let $S^{cis}$ indicate the number of proximal features (cis) and $S^{trans}$=S-$S^{cis}$ the number of distal features (trans). In particular, we here consider a “cis-window” including 50 kb on either side of the DMR. The phenotype matrix, denoted by **Y** , is an N × E matrix while cis and trans standardized genotype matrices, $\mathbf{X}^{cis}$ and $\mathbf{X}^{trans}$, are respectively N × $S^{cis}$ and N × $S^{trans}$. Similarly, $\mathbf{X}^{CMT2}$ denotes the CMT2 genotypes. In the following, □ and □ denote respectively the Kronecker and the element-wise products.

In the model we assume here, the phenotype is written as sum of five contributions

$$\mathcal{L}(\boldsymbol{\mu}, \boldsymbol{\theta}) = \mathcal{N}\left(\mathbf{y} \,|\, \underbrace{\sum_{e=1}^{E} \mu_e^{\text{env}} 1_{\text{env}=e}}_{\text{env. eff.}}, \underbrace{\mathbf{C}^{\text{CMT2}}(\boldsymbol{\theta}) \otimes \mathbf{K}^{\text{CMT2}}}_{\text{CMT2 eff.}} + \underbrace{\mathbf{C}^{cis}(\boldsymbol{\theta}) \otimes \mathbf{K}^{cis}}_{\text{cis genetic eff.}} + \underbrace{\mathbf{C}^{trans}(\boldsymbol{\theta}) \otimes \mathbf{K}^{trans}}_{\text{trans genetic eff.}} + \underbrace{\boldsymbol{\Sigma}(\boldsymbol{\theta}) \otimes \mathbf{I}_N}_{\text{noise}}\right)$$

where **y** = vec**Y**. The mean term contains the environmental component ($\boldsymbol{\mu}^{env}$ is an E vector of fixed effects, and $1_{env=e}$ is a label vector for environment e) while the covariance matrix is sum of four terms: term 1 model genetic contribution from CMT2, terms 2 and 3 model genetic contributions from cis and trans respectively and then there is the noise term. In this context, $\mathbf{K}^{cis}$ and $\mathbf{K}^{trans}$ describe the relationship between individuals due to cis and trans genetic loci respectively and can be calculated from the genotype data as $\mathbf{K}^{cis} = \mathbf{X}^{cis}\mathbf{X}^{cisT}$ and $\mathbf{K}^{trans} = \mathbf{X}^{trans}\mathbf{X}^{transT}$ , similarly for $\mathbf{K}^{CMT2} = \mathbf{X}^{CMT2}\mathbf{X}^{CMT2T}$(they are then normalized such that mean(diag($K^x$)) = 1). On the other hand, $\mathbf{C}^{CMT2}$, $\mathbf{C}^{cis}$, $\mathbf{C}^{trans}$ and **Σ** describe phenotypic correlations across environments due to cis-genetic, trans-genetic and non-genetic contributions, they are unknown and need to be estimated from the data. We indicated with **θ** the parameters of those matrices collectively and with **μ** as alias for $\boldsymbol{\mu}^{env}$.

Parameter inference of **μ** and **θ** is done using maximum likelihood using the LBFGS optimizer. Optimality is reached when $\nabla\mathcal{L}(\boldsymbol{\mu}, \boldsymbol{\theta}) \approx \mathbf{0}$. In particular, we used as condition for optimality. Optimizations were performed DMR-by-DMR using 10 random restarts and stopping as soon as the optimality condition has been met. Only DMRs for which we find a local minimum within the 10 restarts are considered.

In the following, we describe how variance components are calculated once **μ**, $\mathbf{C}^{CMT2}$, $\mathbf{C}^{cis}$, $\mathbf{C}^{trans}$ and **Σ** have been estimated. The phenotypic variability due purely to environment are obtained by the fixed effect considering var($\boldsymbol{\mu}^{env}$). Environment-persistent and environment-specific effects due to x-genetic effects can be dissected by decomposing $C^x$ into the biggest semi-definite positive matrix contained, $a^2\mathbf{1}_{EE}$, and the residual. While $a^2$ will measure the variance explained by x-genetic effects that persist across environments, the diagonal of the residual will measure GxE effects in the two environments. Variance components are averaged across environments and then normalized so that they sum to 1.

# References

Akalin, A., Kormaksson, M., Li, S., Garrett-Bakelman, F.E., Figueroa, M.E., Melnick, A., and Mason, C.E. (2012). methylKit: a comprehensive R package for the analysis of genome-wide DNA methylation profiles. Genome Biol *13*, R87.

Anders, S., and Huber, W. (2010). Differential expression analysis for sequence count data. Genome Biol *11*, R106.

Atwell, S., Huang, Y.S., Vilhjalmsson, B.J., Willems, G., Horton, M., Li, Y., Meng, D., Platt, A., Tarone, A.M., Hu, T.T.*, et al.* (2010). Genome-wide association study of 107 phenotypes in Arabidopsis thaliana inbred lines. Nature *465*, 627-631.

Becker, C., Hagmann, J., Muller, J., Koenig, D., Stegle, O., Borgwardt, K., and Weigel, D. (2011). Spontaneous epigenetic variation in the Arabidopsis thaliana methylome. Nature *480*, 245-249.

Cavrak, V.V., Lettner, N., Jamge, S., Kosarewicz, A., Bayer, L.M., and Mittelsten Scheid, O. (2014). How a retrotransposon exploits the plant's heat stress response for its activation. PLoS Genet *10*, e1004115.

Daxinger, L., Kanno, T., Bucher, E., van der Winden, J., Naumann, U., Matzke, A.J., and Matzke, M. (2009). A stepwise pathway for biogenesis of 24-nt secondary siRNAs and spreading of DNA methylation. Embo J *28*, 48-57.

Dinh, H.Q., Dubin, M., Sedlazeck, F.J., Lettner, N., Mittelsten Scheid, O., and von Haeseler, A. (2012). Advanced methylome analysis after bisulfite deep sequencing: an example in Arabidopsis. PLoS One *7*, e41528.

Eichten, S.R., Briskine, R., Song, J., Li, Q., Swanson-Wagner, R., Hermanson, P.J., Waters, A.J., Starr, E., West, P.T., Tiffin, P.*, et al.* (2013). Epigenetic and genetic influences on DNA methylation variation in maize populations. Plant Cell *25*, 2783-2797.

Finnegan, E.J., Genger, R.K., Peacock, W.J., and Dennis, E.S. (1998). DNA METHYLATION IN PLANTS. Annu Rev Plant Physiol Plant Mol Biol *49*, 223-247.

Gutierrez-Arcelus, M., Lappalainen, T., Montgomery, S.B., Buil, A., Ongen, H., Yurovsky, A., Bryois, J., Giger, T., Romano, L., Planchon, A.*, et al.* (2013). Passive and active DNA methylation and the interplay with genetic variation in gene regulation. Elife *2*, e00523.

Hancock, A.M., Brachi, B., Faure, N., Horton, M.W., Jarymowycz, L.B., Sperone, F.G., Toomajian, C., Roux, F., and Bergelson, J. (2011). Adaptation to climate across the Arabidopsis thaliana genome. Science *334*, 83-86.

Hansen, K.D., Langmead, B., and Irizarry, R.A. (2012). BSmooth: from whole genome bisulfite sequencing reads to differentially methylated regions. Genome Biol *13*, R83.

Henikoff, S., and Comai, L. (1998). A DNA methyltransferase homolog with a chromodomain exists in multiple polymorphic forms in Arabidopsis. Genetics *149*, 307-318.

Ito, H., Gaubert, H., Bucher, E., Mirouze, M., Vaillant, I., and Paszkowski, J. (2011). An siRNA pathway prevents transgenerational retrotransposition in plants subjected to stress. Nature *472*, 115-119.

Jean, G., Kahles, A., Sreedharan, V.T., De Bona, F., and Ratsch, G. (2010). RNA-Seq read alignments with PALMapper. Curr Protoc Bioinformatics *Chapter 11*, Unit 11 16.

Kang, H.M., Sul, J.H., Service, S.K., Zaitlen, N.A., Kong, S.Y., Freimer, N.B., Sabatti, C., and Eskin, E. (2010). Variance component model to account for sample structure in genome-wide association studies. Nat Genet *42*, 348-354.

Kim, S., Plagnol, V., Hu, T.T., Toomajian, C., Clark, R.M., Ossowski, S., Ecker, J.R., Weigel, D., and Nordborg, M. (2007). Recombination and linkage disequilibrium in Arabidopsis thaliana. Nat Genet *39*, 1151-1155.

Li, Q., Eichten, S.R., Hermanson, P.J., and Springer, N.M. (2014). Inheritance patterns and stability of DNA methylation variation in maize near-isogenic lines. Genetics *196*, 667-676.

Lindroth, A.M., Cao, X., Jackson, J.P., Zilberman, D., McCallum, C.M., Henikoff, S., and Jacobsen, S.E. (2001). Requirement of CHROMOMETHYLASE3 for maintenance of CpXpG methylation. Science *292*, 2077-2080.

Liu, Y., Siegmund, K.D., Laird, P.W., and Berman, B.P. (2012). Bis-SNP: Combined DNA methylation and SNP calling for Bisulfite-seq data. Genome Biol *13*, R61.

Long, Q., Rabanal, F.A., Meng, D., Huber, C.D., Farlow, A., Platzer, A., Zhang, Q., Vilhjalmsson, B.J., Korte, A., Nizhynska, V.*, et al.* (2013). Massive genomic variation and strong selection in Arabidopsis thaliana lines from Sweden. Nat Genet *45*, 884-890.

Pecinka, A., Dinh, H.Q., Baubec, T., Rosa, M., Lettner, N., and Mittelsten Scheid, O. (2010). Epigenetic regulation of repetitive elements is attenuated by prolonged heat stress in Arabidopsis. Plant Cell *22*, 3118-3129.

Romon, M., Soustre-Gacougnolle, I., Schmitt, C., Perrin, M., Burdloff, Y., Chevalier, E., Mutterer, J., Himber, C., Zervudacki, J., Montavon, T.*, et al.* (2013). RNA silencing is resistant to low-temperature in grapevine. PLoS One *8*, e82652.

Schmitz, R.J., Schultz, M.D., Urich, M.A., Nery, J.R., Pelizzola, M., Libiger, O., Alix, A., McCosh, R.B., Chen, H., Schork, N.J.*, et al.* (2013). Patterns of population epigenomic diversity. Nature *495*, 193-198.

Sedlazeck, F.J., Rescheneder, P., and von Haeseler, A. (2013). NextGenMap: fast and accurate read mapping in highly polymorphic genomes. Bioinformatics *29*, 2790-2791.

Segura, V., Vilhjalmsson, B.J., Platt, A., Korte, A., Seren, U., Long, Q., and Nordborg, M. (2012). An efficient multi-locus mixed-model approach for genome-wide association studies in structured populations. Nat Genet *44*, 825-830.

Shen, X., Forsberg, S., Pettersson, M., Sheng, Z., and Carlborg, O. (2013). Natural CMT2 variation is associated with genome-wide methylation changes and temperature adaptation. In ArXiv e-prints, pp. 4522.

Shin, J.H., Blay, S., McNeney, B., and Graham, J. (2006). LDheatmap: An R Function for Graphical Display of Pairwise Linkage Disequilibria Between Single Nucleotide Polymorphisms. J Stat Soft *16*.

Statham, A.L., Strbenac, D., Coolen, M.W., Stirzaker, C., Clark, S.J., and Robinson, M.D. (2010). Repitools: an R package for the analysis of enrichment-based epigenomic data. Bioinformatics *26*, 1662-1663.

Stroud, H., Do, T., Du, J., Zhong, X., Feng, S., Johnson, L., Patel, D.J., and Jacobsen, S.E. (2014). Non-CG methylation patterns shape the epigenetic landscape in Arabidopsis. Nat Struct Mol Biol *21*, 64-72.

Stroud, H., Greenberg, M.V., Feng, S., Bernatavichute, Y.V., and Jacobsen, S.E. (2013). Comprehensive analysis of silencing mutants reveals complex regulation of the Arabidopsis methylome. Cell *152*, 352-364.

Takuno, S., and Gaut, B.S. (2012). Body-methylated genes in Arabidopsis thaliana are functionally important and evolve slowly. Mol Biol Evol *29*, 219-227.

Vaughn, M.W., Tanurdzic, M., Lippman, Z., Jiang, H., Carrasquillo, R., Rabinowicz, P.D., Dedhia, N., McCombie, W.R., Agier, N., Bulski, A.*, et al.* (2007). Epigenetic natural variation in Arabidopsis thaliana. PLoS Biol *5*, e174.

Xi, Y., and Li, W. (2009). BSMAP: whole genome bisulfite sequence MAPping program. BMC Bioinformatics *10*, 232.

Zemach, A., Kim, M.Y., Hsieh, P.H., Coleman-Derr, D., Eshed-Williams, L., Thao, K., Harmer, S.L., and Zilberman, D. (2013). The Arabidopsis nucleosome remodeler DDM1 allows DNA methyltransferases to access H1-containing heterochromatin. Cell *153*, 193-205.

Zhao, K., Aranzana, M.J., Kim, S., Lister, C., Shindo, C., Tang, C., Toomajian, C., Zheng, H., Dean, C., Marjoram, P.*, et al.* (2007). An Arabidopsis example of association mapping in structured samples. PLoS Genet *3*, e4.

Zhong, X., Hale, C.J., Law, J.A., Johnson, L.M., Feng, S., Tu, A., and Jacobsen, S.E. (2012). DDR complex facilitates global association of RNA polymerase V to promoters and evolutionarily young transposons. Nat Struct Mol Biol *19*, 870-875.

Zhu, L.J., Gazin, C., Lawson, N.D., Pages, H., Lin, S.M., Lapointe, D.S., and Green, M.R. (2010). ChIPpeakAnno: a Bioconductor package to annotate ChIP-seq and ChIP-chip data. BMC Bioinformatics *11*, 237.

Zilberman, D., Gehring, M., Tran, R.K., Ballinger, T., and Henikoff, S. (2007). Genome-wide analysis of Arabidopsis thaliana DNA methylation uncovers an interdependence between methylation and transcription. Nat Genet *39*, 61-69.

**Acknowledgements** This work was supported multiple sources: the National Human Genome Research Institute of the US National Institutes of Health (P50HG002790 to MN and RMC; PI S. Tavaré); the European Research Council (268962 MAXMAP to MN, Marie Curie FP7 fellowship 253524 to OS); the European Community Framework Programme 7 (283496 transPLANT to MN); the National Institutes of Health Genetics (Training Grant GM07464) to EJO; the Austrian Science Fund (FWF M1369) to MJD; as well as core funding at the GMI. Sequencing was carried out by the Epigenome Center at the University of Southern California, BGI, and the Campus Support Facilities at the Vienna Biocenter. The authors wish to thank: Andrew Smith for preliminary analyses and effectively supervising the project during the transition of MN from USC to GMI; J. Bergelson for providing seed; many members of MN's lab for discussion (in particular D. Filiault, F. Rabanal, P. Novikova, E. Kerdaffrec and Ü. Seren for help with various analysis tasks); A. Hancock and C. Huber for discussions about selection; F. Sedlazeck, and P. Rescheneder for advice on alignment; and F. Berger and O. Mittelsten Scheid for discussions and comments on the paper.

**Author contributions** RMC and MN designed and supervised the study. MSR and PZ generated the raw data and carried out preliminary analysis together with EJO and QS. PD, AK, BV, OS and GR analysed the mRNA-seq data. FPC, DM and OS carried out the variance-component analysis. QL provided pre-publication access to genome sequencing data. VV, JJ, SI and MJD generated follow-up data. MJD carried out most other analyses and wrote the paper together with MN.

**Author information** RNA sequencing and bisulfite sequencing data are available from the Gene Expression Omnibus (GSE54292 and GSE54680). Reprints and permisions information are avalible from nature.com/reprints. The authors declare no competing financial interests.Correspondence and requests for materials should be addressed to MN (magnus.nordborg@gmi.oeaw.ac.at).

# Figures

**Figure 1**

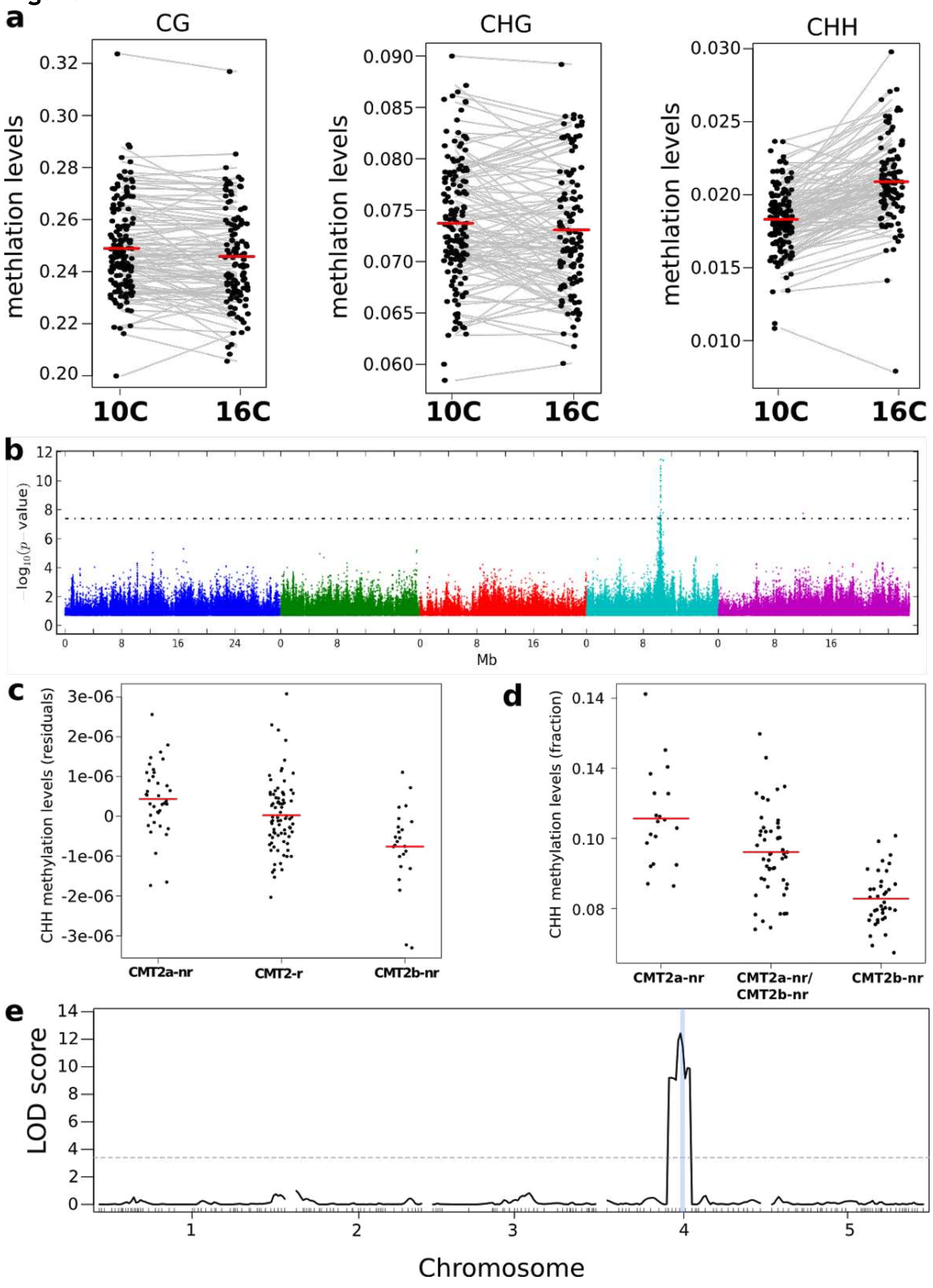

**Figure 1 | The effect of CMT2 on genome-wide CHH methylation levels. a,** Genome-wide average methylation level reaction norms for each accession (156 samples at 10C and 125 samples at 16C). Only CHH levels differ significantly between temperatures (Wilcoxon rank sum test; p = 1.7e-16). **b,** Manhattan plot of GWAS results using average levels of CHH methylation for 151 accessions at 10 C on large transposons as the phenotype (the peak is also seen at 16 C [not shown]). **c,** CHH methylation on large (over 2kb) transposons at 10C by CMT2 two-locus genotype (population sizes are 36, 82, and 24 for CMT2A-nr, CMT2-r and CMTB-nr respectively). The values plotted are the Best Linear Unbiased Predictor (BLUP) estimates after correcting for population structure. Since accessions are homozygous, only four genotypes are possible, of which only three exist due to complete linkage disequilibrium between *CMT2a* and *CMT2b*. **d,**CHH methylation on large transposons by CMT2 genotype in an F2 population of 109 individuals(population sizes are 19, 52, and 38 for CMT2A-nr, CMT2-r and CMTB-nr respectively). **e,** Mapping of CHH methylation of long TEs in the same population. The dotted line indicates a LOD threshold with a genome-wide p-value of 0.05 obtained using 1000 permutations, and the vertical blue line shows the marker interval that contains CMT2.

# Figure 2

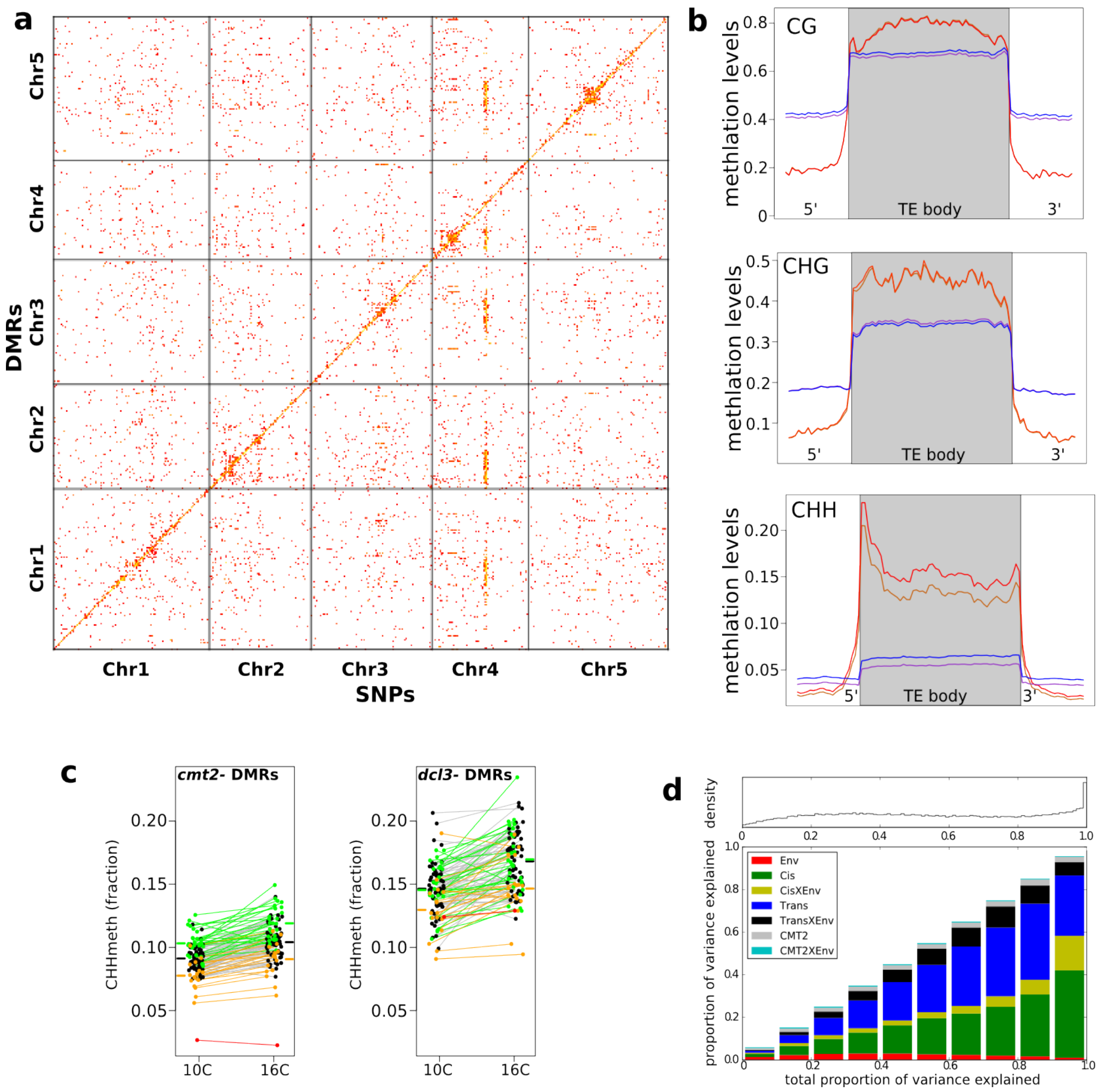

**Figure 2 | Genetic basis CHH methylation variation. a,** GWAS for CHH DMRs at 10C in 151 accessions, defined using 200 bp sliding windows across the genome and selecting the 200,000 most variable ones. For each DMR, SNPs significantly associated at the Bonferroni-corrected 0.05-level are plotted **b,** Average methylation levels over variable transposons at 10 C (orange) vs 16 C (red), and over non-variable transposons at 10 C (purple) vs 16 C (dark blue). Methylation for variable TEs is significantly higher (permutation p-value for CHH methylation = 0.05). **c,** CHH methylation at CMT2- and DCL3-dependent DMRs in natural accessions grown at 10C and 16C (cf. Fig. 1a, each population has 110 individuals). The difference between temperatures was highly significant for both types of DMR (Wilcoxon p-value = 9.1e-11 and p-value = 5.9e-12 respectively). Black points/grey lines indicate accessions with the *CMT2* reference allele; green, the *CMT2a* non-reference allele; and orange, the *CMT2b* non-reference allele. Red is the TAA-03 accession, which has a putative null allele of CMT2. Average methylation levels for each of the genotypes are shown in bars to the side **d,** Variance-components analysis of the CHH DMRs. For each DMR, a mixed model with *cis*, *CMT2*, and genome-wide *trans* effects, plus environment and genetic interactions with environment was fitted (see Methods). DMRs were binned by the total variance explained by the model. The density of DMRs in each bin is shown at the top, and the bottom shows the average variance-decomposition for each bin.

## Figure 3

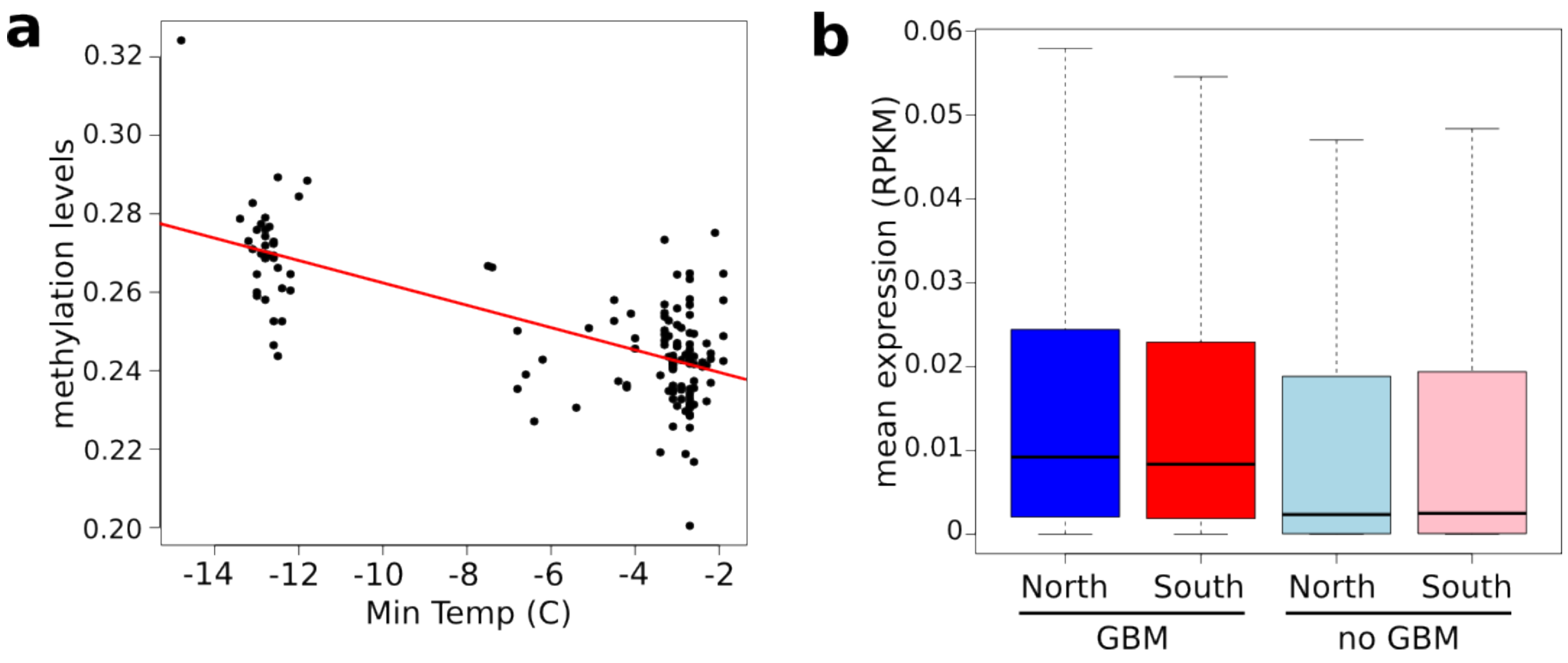

**Figure 3 | Latitudinal difference in GBM and gene expression**. **a,** Global CG methylation levels at 10 C for 151 accessions are strongly correlated with minimum temperature at the location of origin. Results for 16 C are similar. **b,** Genes with GBM are more highly expressed at 10 C in northern (blue) than in southern (red) accessions (wilcoxon rank sum test *p* = 2.1e-03), whereas genes without GBM show the opposite trend (*p* = 1.9e-02). At 16 C the difference for genes with GBM is more significant (p = 6.4e-05), whereas the difference for genes without GBM is insignificant (p = 0.49).

# Figure 4

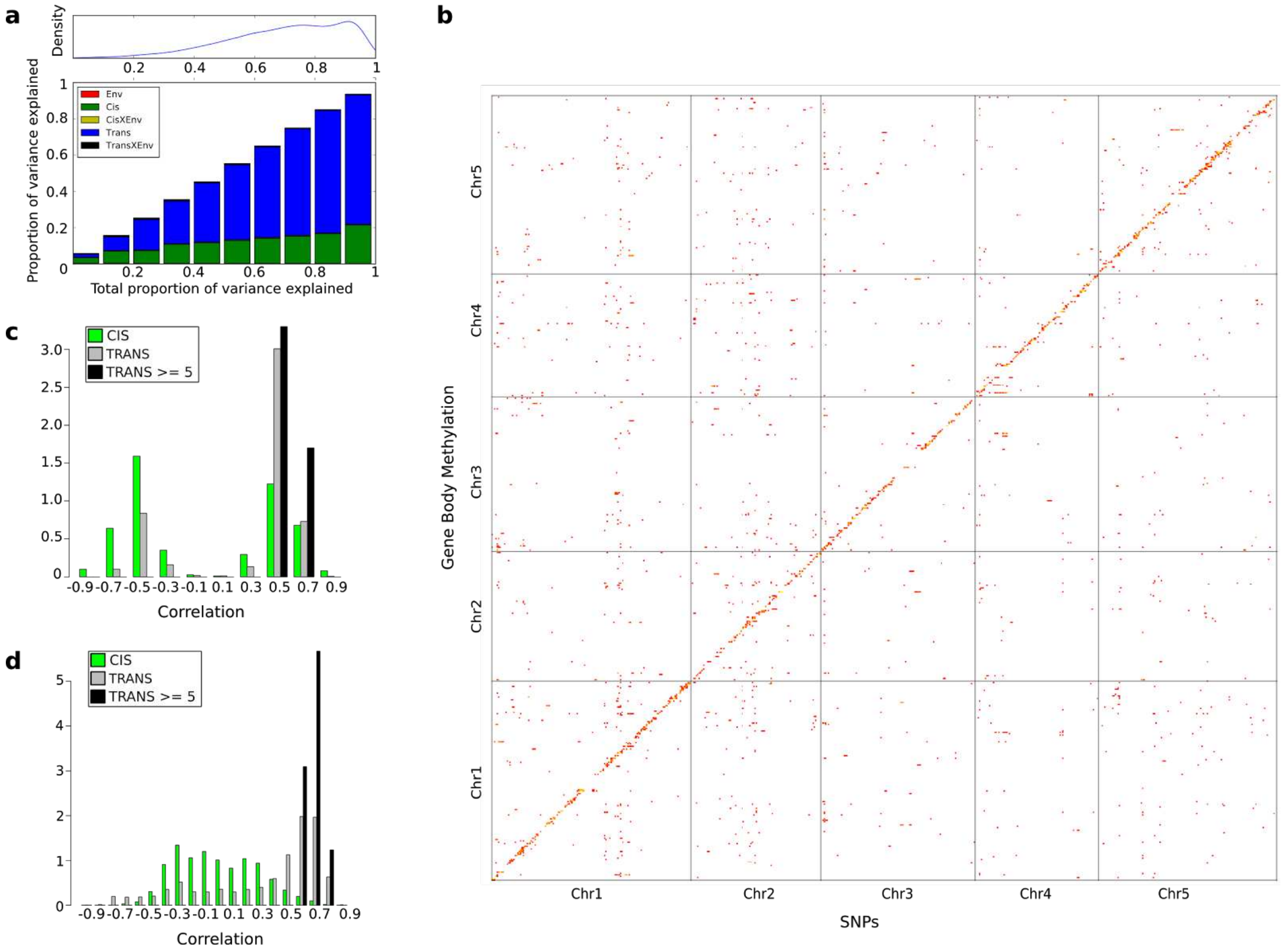

**Figure 4 | The genetic basis of GBM. a,** Variance component analysis of GBM. **b,** Significant associations (Bonferroni-corrected 0.05-level) from a GWAS of GBM for individual genes. **c,** Correlation between non-reference allele at associated SNPs and GBM. **d,** Correlation between non-reference allele at associated SNPs and latitude.

## Supplementary Figures and Tables

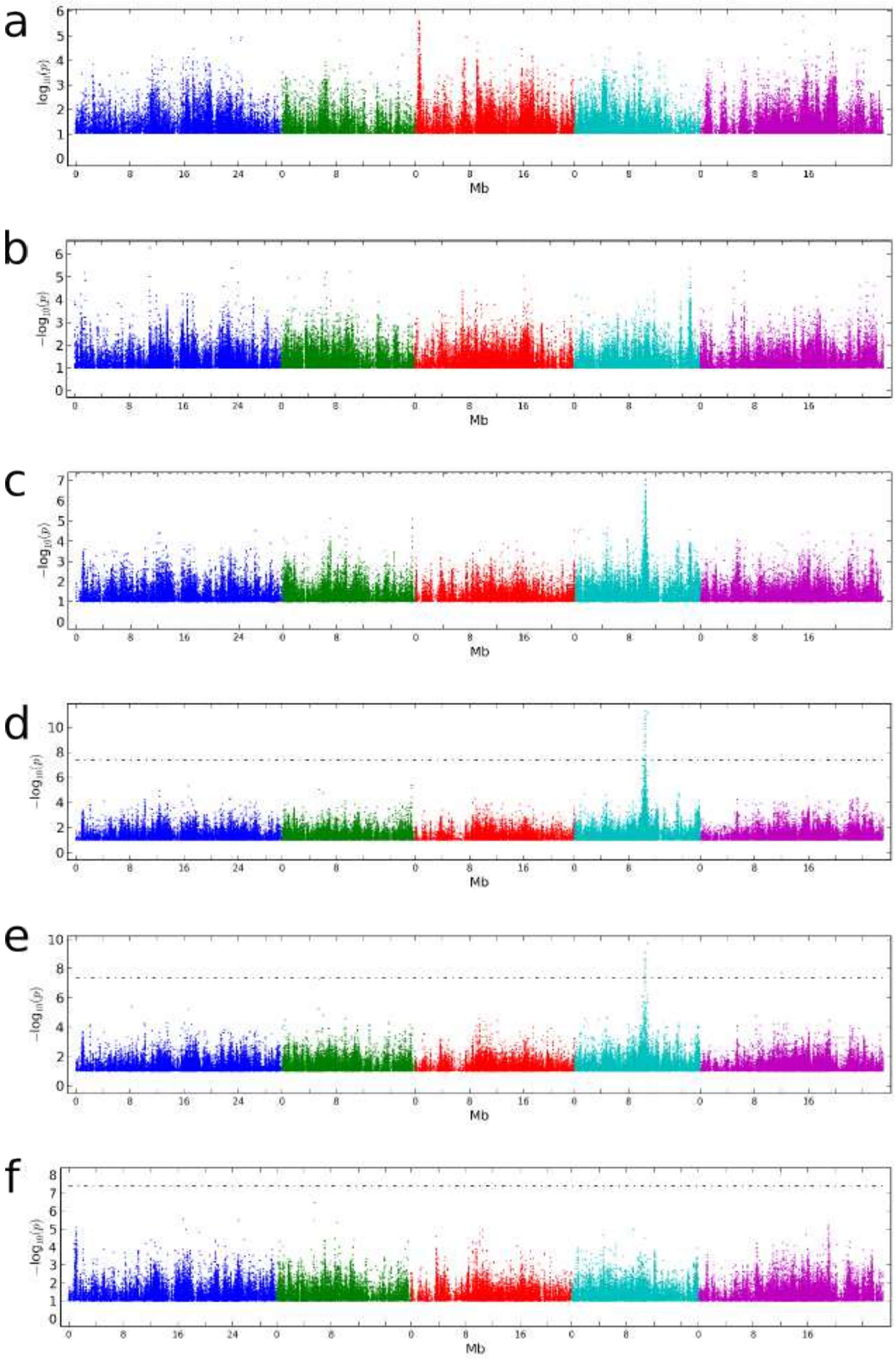

**Supplementary Figure S1 | Manhattan plots of GWAS results for global methylation averages. a**, CG methylation at 10C. **b,** CHG methylation at 10C. **c,** CHH methylation at 10C. Results for methylation at 16C were similar (data not shown). **d,** Stepwise GWAS using average CHH methylation of long TE's as a phenotype without a cofactor. **e,** Including SNP on chr 4 at position 10,459,127 (*CMT2a*) as a cofactor. **f,** Including snps on chr 4 at 10,459,127 (*CMT2a*) and 10,454,628 (*CMT2b*) as cofactors.

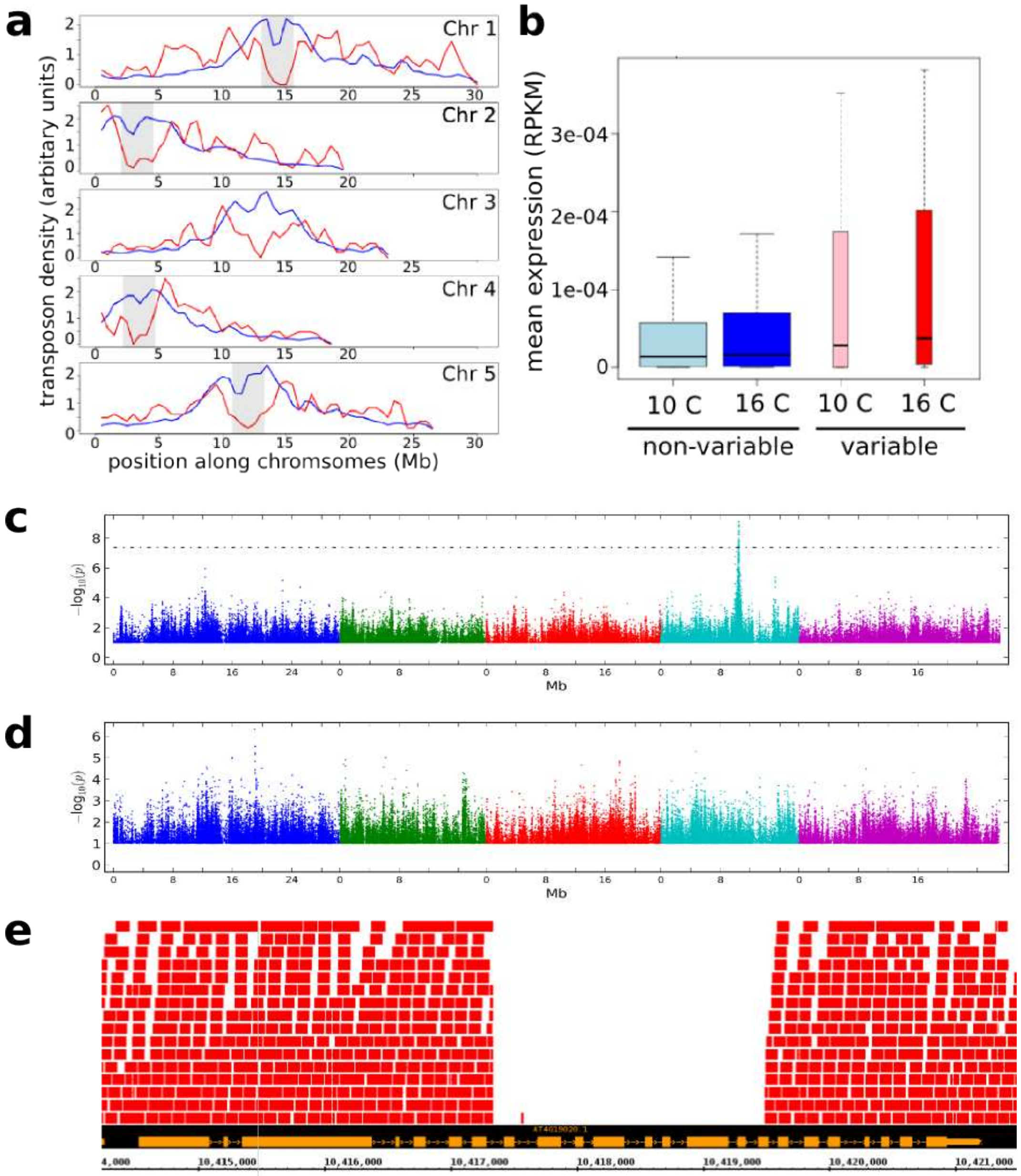

**Supplementary Figure S2 | Features of "variable" TEs and CMT2. a**, The density of variable (red) and non-variable TEs along chromosomes in 500 kb windows. Density is defined as the percentage of the total number in either category in each window, pericentromeric regions are shaded grey. **b**, The expression of TEs at both temperatures. Variable TEs are more highly expressed than non-variable TEs, but the difference is only statistically significant at 16C (Wilcoxon: 10 C, $p = 0.15$; 16 C, $p = 0.023$). **c**, GWAS for CMT2-dependent DMRs at 10C. **d**, GWAS on DCL3-dependent DMRs at 10C. Results from 16C were similar in both cases. **e**, screenshot from a genome browser indicating the lack of read coverage for CMT2 stretching from intron 7 to exon 16 in the accession TAA-03.

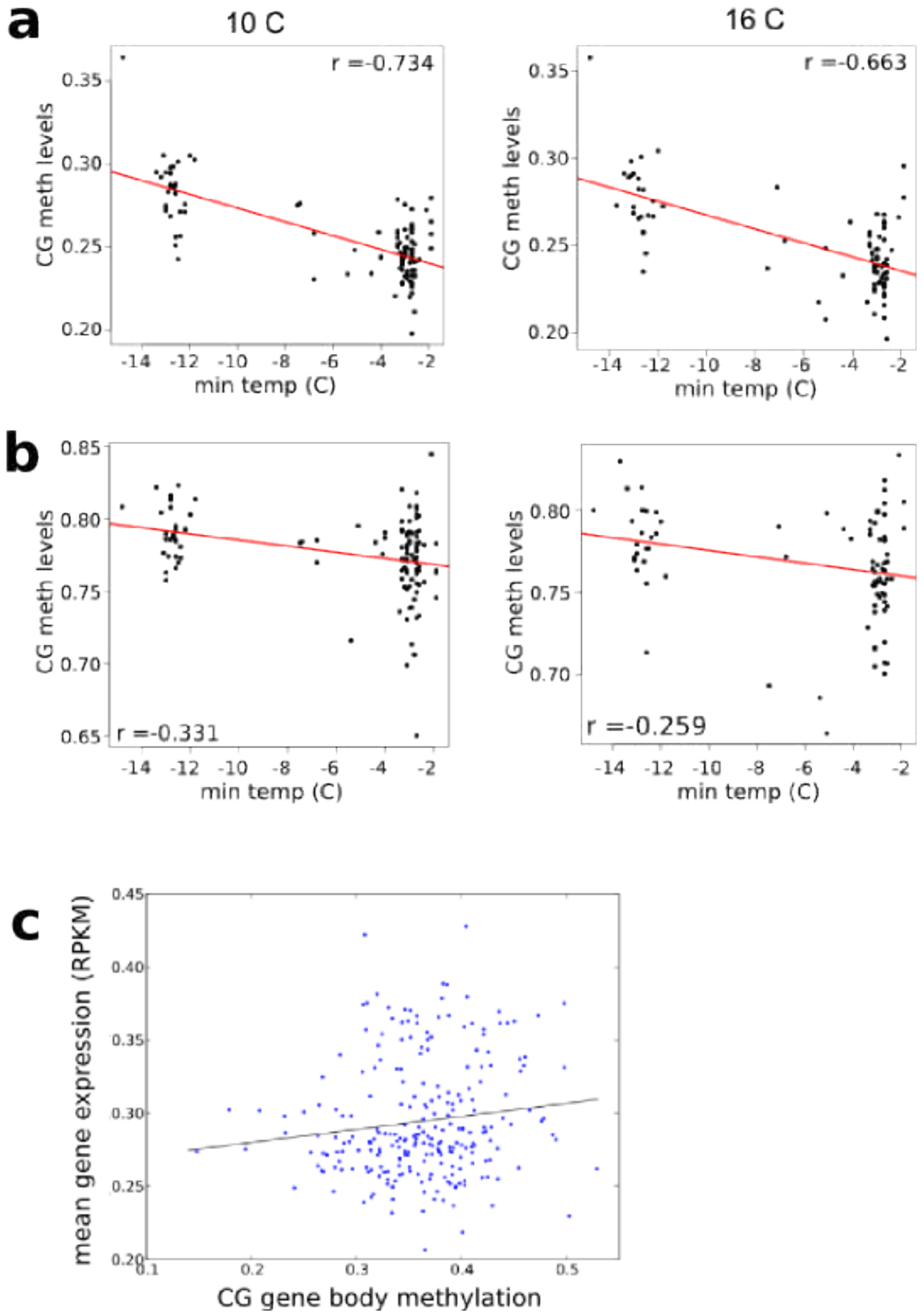

**Supplementary Figure S3 | Correlation between CG methylation levels and the minimum temperature at location of origin. a,** GBM at 10C and 16C. **b**, TE CG methylation at 10C and 16C. **c**, Accessions with higher average GBM tend to have higher average expression (of genes with GBM, normalized by genes without GBM; r = 0.131, p = 0.0386).

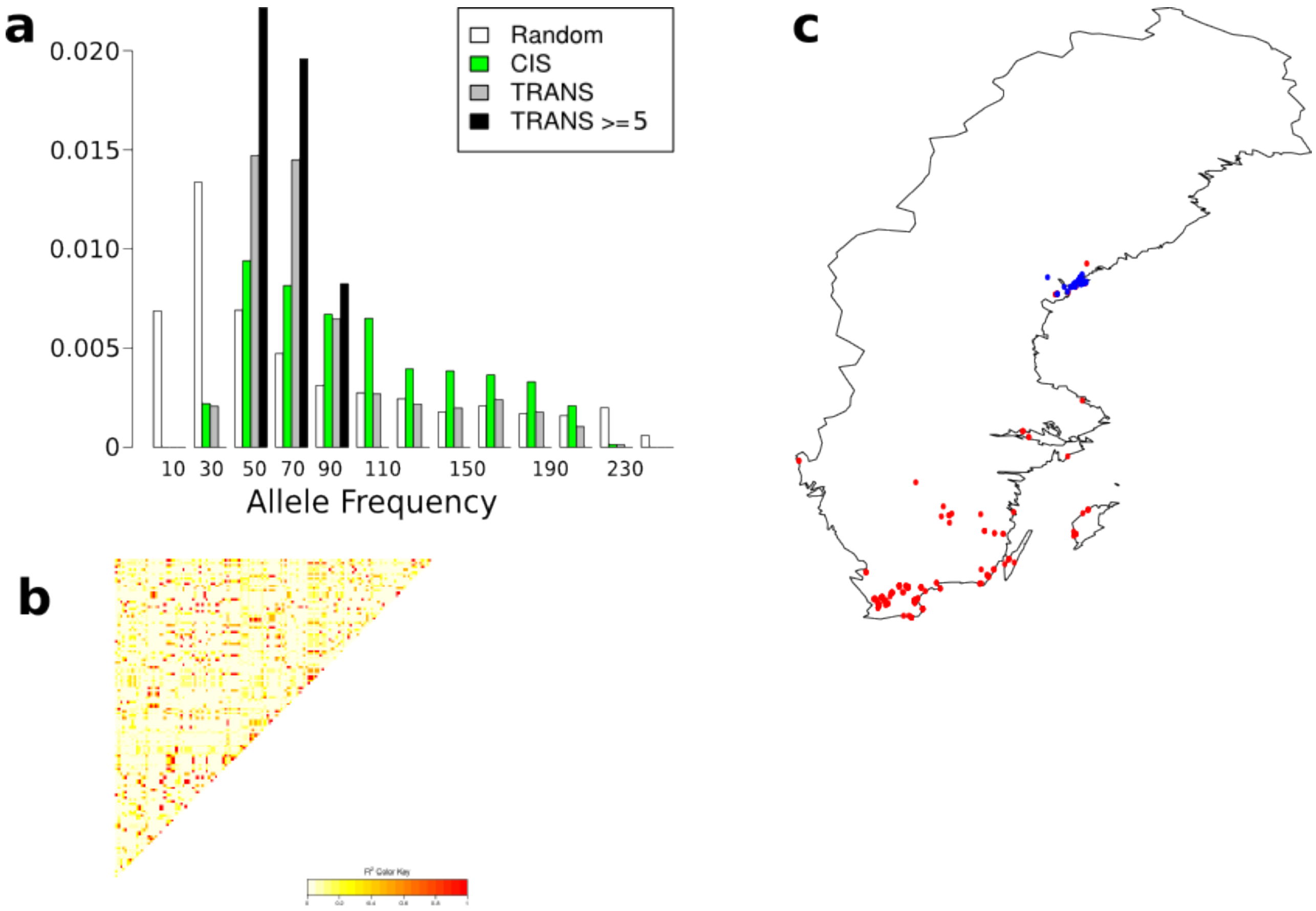

**Supplementary Figure S4 | Distribution of trans-SNPs. a,** Non-reference allele frequency distribution for *cis* and *trans* SNPs compared to random SNPs. **b,** Linkage disequilibrium between the 15 highly associated trans-SNPs. **c,** Accessions carrying the non-reference alleles are limited to northern Sweden (accessions with the non-reference allele at 8 or more of the 15 loci blue, remaining accessions are red).

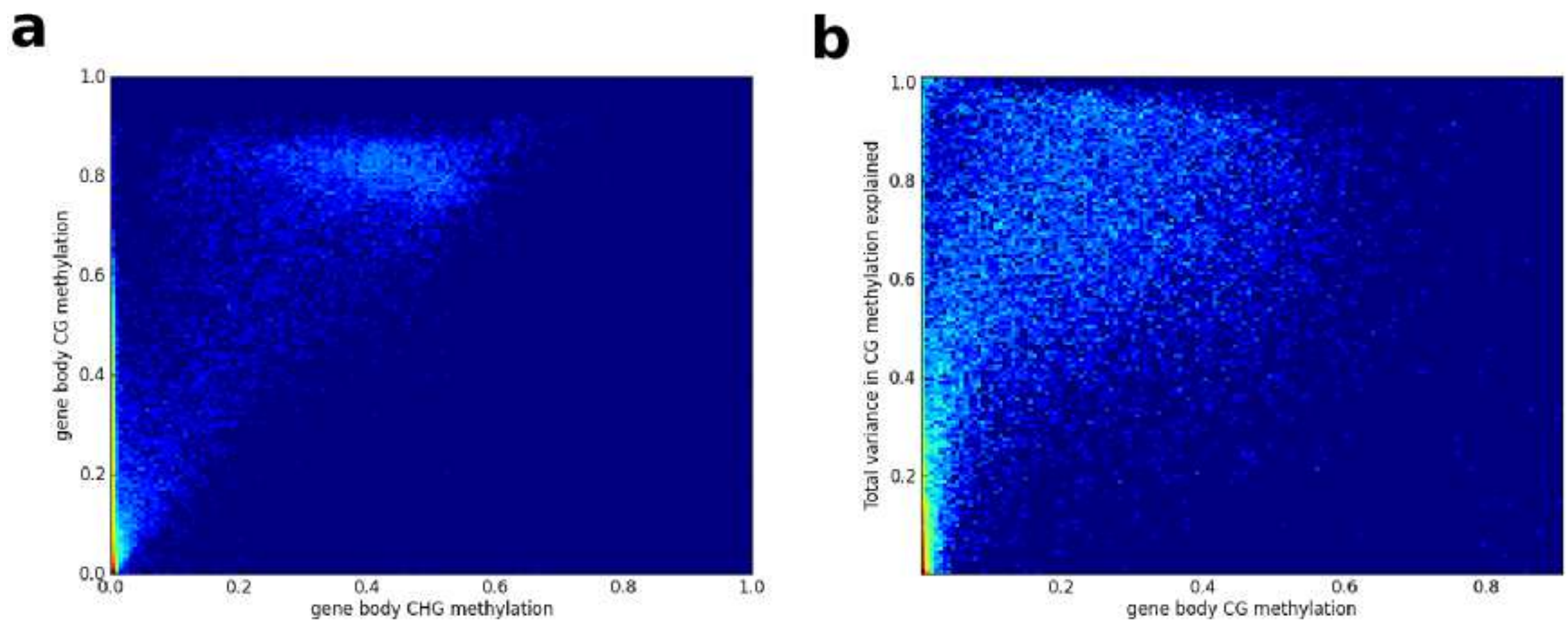

**Supplementary Figure S5 | Filtering of GBM variation data. a,** Genes with low or no CHG methylation have variable levels of CG methylation, while genes with appreciable CHG methylation have very high CG (and CHH) methylation. **b,** Among genes with only CG gene body methylation, variance-component analysis reveals a bimodal distribution of the total variance explained: variation in methylation for genes with low levels of methylation typically does not appear to have a genetic basis.

**Supplementary Table 1 | Super-families (italics) and families that are over-represented among "variable" TEs.**

| TE (*super*-)family | expected | observed | enrichment | 95th quantile |
|---|---|---|---|---|
| *RathE1_cons* | *5* | *26* | *4.56* | *10* |
| *RathE3_cons* | *2* | *9* | *3.23* | *6* |
| *RathE2_cons* | *1* | *5* | *2.52* | *4* |
| *SINE* | *3* | *7* | *2.00* | *7* |
| *RC/Helitron* | *346* | *444* | *1.28* | *368* |
| *DNA/MuDR* | *144* | *184* | *1.27* | *162* |
| ATREP2 | 4 | 53 | 12.07 | 8 |
| RP1_AT | 2 | 27 | 11.59 | 5 |
| ATTIRX1C | 1 | 12 | 11.49 | 3 |
| ATREP13 | 2 | 28 | 9.87 | 6 |
| VANDAL22 | 1 | 11 | 8.56 | 3 |
| SIMPLEHAT1 | 1 | 11 | 7.34 | 4 |
| VANDAL2N1 | 1 | 10 | 7.32 | 3 |
| ATREP8 | 2 | 13 | 6.47 | 5 |
| VANDAL2 | 1 | 7 | 6.08 | 3 |
| ATREP10 | 1 | 10 | 5.93 | 4 |
| AT9NMU1 | 1 | 7 | 5.81 | 3 |
| ATN9_1 | 1 | 10 | 5.75 | 4 |
| SIMPLEHAT2 | 1 | 11 | 5.63 | 4 |
| META1 | 3 | 20 | 5.41 | 7 |
| ATDNAI27T9A | 3 | 15 | 4.83 | 6 |
| ATREP2A | 3 | 15 | 4.83 | 6 |
| ATCOPIA78 | 0 | 3 | 4.67 | 2 |
| VANDAL18NA | 0 | 3 | 4.67 | 2 |
| RathE1_cons | 5 | 26 | 4.56 | 10 |
| VANDAL14 | 0 | 3 | 4.48 | 2 |
| SIMPLEGUY1 | 3 | 13 | 4.19 | 6 |
| ATDNATA1 | 0 | 3 | 4.00 | 2 |
| TNAT2A | 1 | 4 | 3.93 | 3 |
| ATREP7 | 4 | 16 | 3.64 | 8 |
| RathE3_cons | 2 | 9 | 3.23 | 6 |
| ATREP14 | 1 | 4 | 3.18 | 3 |
| ATREP16 | 1 | 4 | 3.18 | 3 |
| LIMPET1 | 3 | 9 | 2.90 | 6 |
| ATREP6 | 4 | 14 | 2.89 | 8 |
| ARNOLDY2 | 7 | 22 | 2.85 | 13 |
| ATSINE4 | 2 | 7 | 2.67 | 5 |
| ATDNAI27T9C | 2 | 7 | 2.42 | 6 |
| ATREP3 | 38 | 92 | 2.39 | 49 |
| ARNOLDY1 | 6 | 14 | 2.21 | 11 |
| ATREP1 | 13 | 23 | 1.73 | 19 |
| HELITRONY3 | 37 | 51 | 1.36 | 48 |

**Supplementary Table 2 | Correlation between methylation levels and environment-of-origin variables (Hancock et al., 2011).** r = Pearson's correlation, rho = Spearman's rank correlation, p-value = significance of rho.

| Environmental variable | Growing temp. | CG | | | CHG | | | CHH | | |
|---|---|---|---|---|---|---|---|---|---|---|
| | | r | rho | p-value | r | rho | p-value | r | rho | p-value |
| Latitude | 10 | 0.69 | 0.52 | 7.8E-11 | -0.24 | -0.19 | 2.7E-02 | 0.10 | 0.14 | 1.1E-01 |
| | 16 | 0.62 | 0.47 | 3.2E-07 | -0.21 | -0.20 | 4.2E-02 | 0.04 | -0.11 | 2.5E-01 |
| Longitude | 10 | 0.59 | 0.54 | 1.2E-11 | -0.14 | -0.09 | 3.1E-01 | 0.23 | 0.28 | 7.5E-04 |
| | 16 | 0.55 | 0.53 | 4.4E-09 | -0.12 | -0.03 | 7.4E-01 | 0.14 | 0.15 | 1.2E-01 |
| Temperature seasonality | 10 | 0.68 | 0.49 | 1.6E-09 | -0.27 | -0.24 | 4.8E-03 | 0.09 | 0.09 | 2.8E-01 |
| | 16 | 0.62 | 0.42 | 1.1E-05 | -0.23 | -0.26 | 6.6E-03 | 0.04 | -0.12 | 2.1E-01 |
| Max. temp. (warmest month) | 10 | -0.14 | 0.06 | 4.6E-01 | -0.07 | -0.13 | 1.3E-01 | 0.14 | 0.20 | 2.0E-02 |
| | 16 | -0.03 | 0.10 | 2.9E-01 | -0.10 | -0.20 | 3.8E-02 | 0.05 | 0.03 | 7.3E-01 |
| Min. temp. (coldest month) | 10 | -0.70 | -0.56 | 9.1E-13 | 0.27 | 0.21 | 1.2E-02 | -0.07 | -0.06 | 4.7E-01 |
| | 16 | -0.63 | -0.48 | 2.7E-07 | 0.24 | 0.24 | 1.4E-02 | 0.00 | 0.19 | 5.6E-02 |
| Precipitation (wettest montth) | 10 | 0.45 | 0.52 | 1.2E-10 | -0.25 | -0.27 | 1.2E-03 | -0.20 | -0.12 | 1.7E-01 |
| | 16 | 0.29 | 0.43 | 4.0E-06 | -0.26 | -0.24 | 1.2E-02 | -0.22 | -0.19 | 5.8E-02 |
| Precipitation (driest montth) | 10 | 0.31 | 0.40 | 1.5E-06 | -0.33 | -0.29 | 6.5E-04 | -0.24 | -0.21 | 1.6E-02 |
| | 16 | 0.21 | 0.32 | 7.4E-04 | -0.26 | -0.24 | 1.4E-02 | -0.15 | -0.18 | 6.0E-02 |
| Precipitation seasonality | 10 | 0.42 | 0.44 | 7.1E-08 | -0.07 | -0.16 | 5.4E-02 | 0.05 | 0.01 | 9.0E-01 |
| | 16 | 0.36 | 0.37 | 1.2E-04 | -0.13 | -0.16 | 1.1E-01 | 0.01 | -0.01 | 9.1E-01 |
| PAR (spring) | 10 | 0.04 | 0.22 | 8.9E-03 | 0.20 | 0.18 | 3.7E-02 | 0.24 | 0.23 | 7.3E-03 |
| | 16 | 0.03 | 0.18 | 6.6E-02 | 0.27 | 0.21 | 3.5E-02 | 0.38 | 0.35 | 2.8E-04 |
| Length of growing season | 10 | -0.59 | -0.57 | 5.5E-13 | 0.24 | 0.23 | 7.3E-03 | -0.16 | -0.18 | 3.3E-02 |
| | 16 | -0.58 | -0.54 | 4.0E-09 | 0.23 | 0.21 | 3.0E-02 | -0.04 | 0.01 | 8.9E-01 |
| No. consecutive cold days | 10 | 0.60 | 0.53 | 4.0E-11 | -0.19 | -0.13 | 1.2E-01 | 0.17 | 0.28 | 1.1E-03 |
| | 16 | 0.57 | 0.53 | 4.2E-09 | -0.17 | -0.09 | 3.7E-01 | 0.10 | 0.08 | 4.1E-01 |
| No. consecutive frost-free days | 10 | -0.59 | -0.49 | 1.2E-09 | 0.29 | 0.27 | 1.5E-03 | 0.02 | 0.03 | 7.1E-01 |
| | 16 | -0.51 | -0.39 | 4.9E-05 | 0.30 | 0.30 | 1.6E-03 | 0.07 | 0.13 | 1.9E-01 |
| Relative humidity (spring) | 10 | 0.62 | 0.47 | 5.6E-09 | -0.23 | -0.18 | 3.9E-02 | 0.09 | 0.06 | 4.5E-01 |
| | 16 | 0.53 | 0.37 | 1.2E-04 | -0.20 | -0.26 | 7.6E-03 | 0.04 | -0.08 | 4.3E-01 |
| Daylength (spring) | 10 | 0.69 | 0.50 | 7.2E-10 | -0.27 | -0.21 | 1.4E-02 | 0.08 | 0.05 | 5.7E-01 |
| | 16 | 0.63 | 0.41 | 1.5E-05 | -0.23 | -0.29 | 2.7E-03 | 0.04 | -0.17 | 8.7E-02 |
| Aridity | 10 | 0.53 | 0.49 | 8.4E-10 | -0.35 | -0.31 | 1.9E-04 | -0.18 | -0.21 | 1.3E-02 |
| | 16 | 0.43 | 0.42 | 8.4E-06 | -0.28 | -0.24 | 1.3E-02 | -0.13 | -0.20 | 3.8E-02 |

**Supplementary Table 3 | 15 SNPs associated with GBM at 5 or more genes.**

| Chr | Position | Associated with GBM at how many genes? | Non-reference allele count | SNP -latitude correlation | Overlap with sweep (Long et al., 2013) | Overlap with min. temp. assoc. SNPs (Hancock et al., 2011) |
|---|---|---|---|---|---|---|
| 1 | 912291 | 8 | 42 | 0.73 | none | 1_914088_0.21 |
| 1 | 4405103 | 5 | 66 | 0.64 | none | none |
| 1 | 7614101 | 5 | 48 | 0.66 | none | none |
| 1 | 19755967 | 5 | 88 | 0.75 | none | 1_19757140_0.24 |
| 2 | 6998631 | 6 | 55 | 0.87 | 2_6931030 | none |
| 2 | 7655016 | 6 | 81 | 0.61 | 2_7613651 | none |
| 2 | 7660469 | 9 | 55 | 0.78 | 2_7613651 | 2_7662427_0.30 |
| 2 | 7666059 | 5 | 69 | 0.72 | 2_7613651 | 2_7665507_0.25 |
| 2 | 7680882 | 5 | 82 | 0.63 | 2_7613651 | none |
| 2 | 7915712 | 6 | 51 | 0.83 | none | 2_7913782_0.23 |
| 2 | 9382495 | 5 | 73 | 0.71 | none | 2_9383856_0.34 |
| 2 | 9653878 | 9 | 48 | 0.80 | none | none |
| 3 | 419309 | 8 | 66 | 0.68 | none | none |
| 4 | 519982 | 8 | 66 | 0.70 | none | none |
| 4 | 13290034 | 5 | 74 | 0.74 | none | none |

**Supplementary Table 4 | Genes in the plant chromatin database that are within 100 kb of one of the 15 SNPs associated with GBM at 5 or more genes.**

| ChromDB | Locus |
|---|---|
| ARID3 | AT2G17410 |
| ARP3 | AT1G13180 |
| CHB4 | AT1G21700 |
| CHR9 | AT1G03750 |
| CHR35 | AT2G16390 |
| CONS3 | AT3G02380 |
| DNG12 | AT1G21710 |
| FLCP39 | AT3G02310 |
| FLCP16 | AT2G22630 |
| FLCP9 | AT2G22540 |
| GTI1 | AT2G22720 |
| HMGB4 | AT2G17560 |
| JMJ27 | AT4G00990 |
| NFA1 | AT4G26110 |
| SDG23 | AT2G22740 |
| SDG37 | AT2G17900 |
| YDG2 | AT2G18000 |
| HON3 | AT2G18050 |

**Supplementary Table 5 | Genes within 100 kb of the 15 SNPs associated with GBM at 5 or more genes whose expression is also correlated with the SNP.**

| SNP | Locus | desciption | p-value |
|---|---|---|---|
| 1_19755967 | AT1G53030 | encodes a copper chaperone | 4.72E-07 |
| 1_19755967 | AT1G52880 | **NO APICAL MERISTEM (NAM) Transcription factor with a NAC domain** | 5.47E-07 |
| 1_19755967 | AT1G52990 | thioredoxin family protein | 2.36E-05 |
| 1_19755967 | AT1G52780 | Protein of unknown function (DUF2921) | 1.46E-04 |
| 1_4405103 | AT1G12750 | RHOMBOID-like protein 6 (RBL6); FUNCTIONS IN: serine-type endopeptidase activity; | 3.74E-08 |
| 1_4405103 | AT1G12790 | RuvA domain 2-like | 2.76E-05 |
| 1_4405103 | AT1G12730 | GPI transamidase subunit | 2.81E-05 |
| 1_4405103 | AT1G13080 | CYTOCHROME P450 FAMILY 71 SUBFAMILY B POLYPEPTIDE 2 (CYP71B2) | 1.65E-04 |
| 1_7614101 | AT1G21790 | TRAM LAG1 and CLN8 (TLC) lipid-sensing domain containing protein | 1.10E-05 |
| 1_7614101 | AT1G21900 | Encodes an ER-localized p24 protein | 8.81E-05 |
| 1_7614101 | AT1G21760 | **F-BOX PROTEIN 7 (FBP7) putative translation regulator in temperature stress response** | 8.54E-04 |
| 1_912291 | AT1G03660 | Ankyrin-repeat containing protein | 1.26E-10 |
| 1_912291 | AT1G03770 | **RING1B protein with similarity to polycomb repressive core complex1 (PRC1)** | 5.76E-07 |
| 1_912291 | AT1G03940 | HXXXD-type acyl-transferase family protein | 1.18E-06 |
| 1_912291 | AT1G03610 | Protein of unknown function (DUF789) | 6.91E-06 |
| 1_912291 | AT1G03580 | pseudogene with weak similarity to ubiquitin-specific protease 12 | 1.29E-05 |
| 1_912291 | AT1G03830 | Guanylate-binding family protein | 3.50E-05 |
| 2_6998631 | AT2G16340 | unknown protein | 1.35E-08 |
| 2_6998631 | AT2G16210 | **Transcriptional factor B3 family protein** | 1.69E-04 |
| 2_7666059 | AT2G17630 | Pyridoxal phosphate (PLP)-dependent transferases superfamily protein | 2.47E-18 |
| 2_7660469 | AT2G17620 | Cyclin B2;1 (CYCB2;1) | 9.68E-07 |
| 2_7655016 | AT2G17740 | Cysteine/Histidine-rich C1 domain family protein | 1.22E-04 |
| 2_7655016 | AT2G17420 | NADPH-DEPENDENT THIOREDOXIN REDUCTASE 2 (NTR2) | 9.96E-04 |
| 2_7666059 | AT2G17430 | MILDEW RESISTANCE LOCUS O 7 (MLO7) | 7.56E-04 |
| 2_7915712 | AT2G18100 | Protein of unknown function (DUF726) | 1.73E-06 |
| 2_7915712 | AT2G17980 | ATSLY member of SLY1 Gene Family | 1.33E-05 |
| 2_7915712 | AT2G18400 | ribosomal protein L6 family protein | 1.26E-04 |
| 2_7915712 | AT2G18150 | Haem peroxidase | 8.05E-04 |
| 2_7915712 | AT2G18050 | **HISTONE H1-3 (HIS1-3)** | 9.47E-04 |
| 2_9382495 | AT2G22260 | HOMOLOG OF E. COLI ALKB (ALKBH2) enzyme involved in DNA methylation damage repair | 1.21E-08 |
| 2_9382495 | AT2G21850 | Cysteine/Histidine-rich C1 domain family protein | 5.38E-06 |
| 2_9382495 | AT2G22240 | MYO-INOSITOL-1-PHOSPHATE SYNTHASE 1 (MIPS1) | 8.71E-05 |
| 2_9382495 | AT2G21940 | SHIKIMATE KINASE 1 (ATSK1) localized to the chloroplast | 1.80E-04 |
| 2_9653878 | AT2G22660 | Protein of unknown function (duplicated DUF1399) | 2.22E-14 |
| 2_9653878 | AT2G22900 | Galactosyl transferase GMA12/MNN10 family protein | 5.08E-09 |
| 2_9653878 | AT2G22830 | squalene epoxidase 2 (SQE2) | 3.91E-06 |
| 2_9653878 | AT2G22640 | BRICK1 (BRK1) | 6.17E-05 |
| 2_9653878 | AT2G22540 | **SHORT VEGETATIVE PHASE (SVP) Floral repressor involved in thermosensory pathway** | 2.46E-04 |
| 2_9653878 | AT2G22570 | NICOTINAMIDASE 1 (NIC1) | 2.67E-04 |
| 2_9653878 | AT2G22770 | **NAI1 Transcription factor** | 7.71E-04 |
| 3_419309 | AT3G02220 | Protein of unknown function (DUF2039) | 2.06E-16 |
| 3_419309 | AT3G02230 | REVERSIBLY GLYCOSYLATED POLYPEPTIDE 1 (RGP1) | 4.58E-14 |
| 3_419309 | AT3G02300 | Regulator of chromosome condensation (RCC1) family protein | 1.25E-10 |
| 3_419309 | AT3G02120 | hydroxyproline-rich glycoprotein family protein | 1.81E-09 |
| 3_419309 | AT3G01980 | Short-chain dehydrogenase/reductase (SDR) | 3.91E-09 |
| 3_419309 | AT3G02370 | unknown protein | 4.53E-08 |
| 3_419309 | AT3G02020 | ASPARTATE KINASE 3 (AK3) | 4.18E-07 |
| 3_419309 | AT3G02160 | **Bromodomain transcription factor** | 2.60E-06 |
| 3_419309 | AT3G02390 | unknown chloroplast protein | 5.60E-06 |
| 3_419309 | AT3G02050 | K+ UPTAKE TRANSPORTER 3 (KUP3) | 1.28E-05 |
| 3_419309 | AT3G02125 | unknown chloroplast protein | 2.12E-05 |
| 3_419309 | AT3G02200 | Proteasome component (PCI) domain protein | 1.16E-04 |
| 3_419309 | AT3G02180 | SPIRAL1-LIKE3 Regulates cortical microtubule organization | 4.56E-04 |
| 3_419309 | AT3G02250 | O-fucosyltransferase family protein | 5.31E-04 |
| 3_419309 | AT3G02110 | serine carboxypeptidase-like 25 (scpl25) | 6.18E-04 |
| 4_13290034 | AT4G26255 | **Non-coding RNA of unknown function** | 1.67E-13 |
| 4_13290034 | AT4G26450 | WPP DOMAIN INTERACTING PROTEIN 1 (WIP1) | 1.13E-04 |
| 4_13290034 | AT4G26230 | Ribosomal protein L31e family protein | 1.74E-04 |
| 4_13290034 | AT4G26160 | ATYPICAL CYS HIS RICH THIOREDOXIN 1 (ACHT1) | 5.72E-04 |
| 4_519982 | AT4G01090 | Protein of unknown function (DUF3133) | 1.23E-06 |
| 4_519982 | AT4G01230 | Reticulon family protein | 2.33E-05 |
| 4_519982 | AT4G01410 | Late embryogenesis abundant (LEA) hydroxyproline-rich glycoprotein family | 5.44E-05 |
| 4_519982 | AT4G01330 | Serine/threonine-protein kinase | 2.22E-04 |
| 4_519982 | AT4G01200 | Calcium-dependent lipid-binding (CaLB domain) family protein | 3.93E-04 |
| 4_519982 | AT4G01390 | TRAF-like family protein | 3.99E-04 |
| 4_519982 | AT4G01040 | Glycosyl hydrolase superfamily protein | 5.66E-04 |
| 4_519982 | AT4G01000 | Ubiquitin-like superfamily protein | 8.55E-04 |

**Alignment of CMT2 sequences from different Accessions:**

_ref = reference (Columbia genotype) _altA = CMT2A-nr, _altB= CMT2B-nr

```
TAIR10          MLSPAKCESEEAQAPLDLHSSSRSEPECLSLVLWCPNPEEAAPSSTRELIKLPDNGEMSL 60
992_ref         MLSPAKCESEEAQAPLDLHSSSRSEPECLSLVLWCPNPEEAAPSSTRELIKLPDNGEMSL 60
6918_ref        MLSPAKCESEEAQAPLDLHSSSRSEPECLSLVLWCPNPEEAAPSSTRELIKLPDNGEMSL 60
6188_altA       MLSPAKCESEEAQAPLDLHSSSRSEPECLSLVLWCPNPEEAAPSSTRELIKLPDNGEMSL 60
1061_altA       MLSPAKCESEEAQAPLDLHSSSRSEPECLSLVLWCPNPEEAAPSSTRELIKLPDNGEMSL 60
6113_altB       MLSPAKCESEEAQAPLDLHSSSRSEPECLSLVLWCPNPEEAAPSSTRELIKLPDNGEMSL 60
6191_altB       MLSPAKCESEEAQAPLDLHSSSRSEPECLSLVLWCPNPEEAAPSSTRELIKLPDNGEMSL 60
                ************************************************************

TAIR10          RRSTTLNCNSPEENGGEGRVSQRKSSRGKSQPLLMLTNGCQLRRSPRFRALHANFDNVCS 120
992_ref         RRSTTLNCNSPEENGGEGRVSQRKSSRGKSQPLLMLTNGCQLRRSPRFRAVHANFDNVCS 120
6918_ref        RRSTTLNCNSPEENGGEGRVSQRKSSRGKSQPLLMLTNGCQLRRSPRFRAVHANFDNVCS 120
6188_altA       RRSTTLNCNSPEENGGEGRVSQRKSSRGKSQPLLMLTNGCQLRRSPRFRALHANFDNVCS 120
1061_altA       RRSTTLNCNSPEENGGEGRVSQRKSSRGKSQPLLMLTNGCQLRRSPRFRAVHANFDNVCS 120
6113_altB       RRSTTLNCNSPEENGGEGRVSQRKSSRGKSQPLLMLTNGCQLRRSPRFRAVHANFDNVCS 120
6191_altB       RRSTTLNCNSPEENGGEGRVSQRKSSRGKSQPLLMLTNGCQLRRSPRFRAVHANFDNVCS 120
                **************************************************:*********

TAIR10          VPVTKGGVSQRKFSRGKSQPLLTLTNGCQLRRSPRFRAVDGNFDSVCSVPVTGKFGSRKR 180
992_ref         VPVTEGGVSQRNSSRGKSQPLLTLTNGCQLRRSPRSRAVDGNFDSVCSVPVTGKFGSRKR 180
6918_ref        VPVTEGGVSQRNSSRGKSQPLLTLTNGCQLRRSPRFRAVDGNFDSVCSVPVTGKFGSRKR 180
6188_altA       VPVTKGGVSQRKFSRGKSQPLLTLTNGCQLRRSPRFRAVDGNFDSVCSVPVTGKFGSRKR 180
1061_altA       VPVTEGGVSQRNSSRGKSQPLLTLTNGCQLRRSPRFRAVDGNFDSVCSVPVTGKFGSRKR 180
6113_altB       VPVTEGGVSQRNSSRGKSQPLLTLTNGCQLRRSPRFKAVDGNFDSVCSVPVTGKFGSRKR 180
6191_altB       VPVTEGGVSQRNSSRGKSQPLLTLTNGCQLRRSPRFKAVDGNFDSVCSVPVTGKFGSRKR 180
                ****:******: ********************** :***********************

TAIR10          KSNSALDKKESSDSEGLTFKDIAVIAKSLEMEIISECQYKNNVAEGRSRLQDPAKRKVDS 240
992_ref         KSNSALDKKESSDSEGLTFKDIAVIAKSLEMEIISECQYKNNVAEGRSKLQDPAKRKVDS 240
6918_ref        KSNSALDKKESSDSEGLTFKDIAVIAKSLEMEIISECQYKNNVAEGRSRLQDPAKRKVDS 240
6188_altA       KSNSALDKKESSDSEGLTFKDIAVIAKSLEMEIISECQYKNNVAEGRSRLQDPAKRKVDS 240
1061_altA       KSNSALDKKESSDSEGLTFKDIAVIAKSLEMEIISECQYKNNVAEGRSRLQDPAKRKVDS 240
6113_altB       KSNSALDKKESSDSEGLTFKDIAVIAKSLEMEIISECQYKNNVAEGRSRLQDPAKRKVDS 240
6191_altB       KSNSALDKKESSDSEGLTFKDIAVIAKSLEMEIISECQYKNNVAEGRSRLQDPAKRKVDS 240
                ***********************************************:************

TAIR10          DTLLYSSINSSKQSLGSNKRMRRSQRFMKGTENEGEENLGKSKGKGMSLASCSFRRSTRL 300
992_ref         DTLLYSSINSSKQSLGSNKRMRRSQRFMKGTENEGEENLGKSKGKGMSLASCSFRRSTRL 300
6918_ref        DTLLSSSINSSKQSLGSNKRMRRSQRFMKGTENEGEENLGKSKGKGMSLASCSFRRSTRL 300
6188_altA       DTLSSSSINSSKQNLGSNKRMRRSQRFMKGTENEGVENLGKSKGKGMSLASCSFRRSTRL 300
1061_altA       DTLSSSSINSSKQNLGSNKRMRRSQRFMKGTENEGVENLGKSKGKGMSLASCSFRRSTRL 300
6113_altB       DTLLYSSINSSKQSLGSNKRMRRSQRFMKGTENEGEENLGKSKGKGMSLASCSFRRSTRL 300
6191_altB       DTLLYSSINSSKQSLGSNKRMRRSQRFMKGTENEGEENLGKSKGKGMSLASCSFRRSTRL 300
                ***  ********.********************* ************************

TAIR10          SGTVETGNTETLNRRKDCGPALCGAEQVRGTERLVQISKKDHCCEAMKKCEGDGLVSSKQ 360
992_ref         SGTVETGNTETLNRRKDCGPALCGAEQVRGTERLVQISKNDHCCEAMKKCEGDGLVSSKQ 360
6918_ref        SGTVETGNTETLNRRKDCGPALCGAEQVRGTERLVQISKNDHCCEAMKKCEGDGLVSSKQ 360
6188_altA       SGTVETGNTETLNRRKDCGPALCGAEQVRGTERLVQISKNDHCCEAMKKCEGDGLVSSKQ 360
1061_altA       SGTVETGNTETLNRRKDCGPALCGAEQVRGTERLVQISKNDHCCEAMKKCEGDGLVSSKQ 360
6113_altB       SGTVETGNTETLNRRKDCGPALCGAEQVRGTERLVQISKNDHCCEAMKKCEGDGLVSSKQ 360
6191_altB       SGTVETGNTETLNRRKDCGPALCGAEQVRGTERLVQISKNDHCCEAMKKCEGDGLVSSKQ 360
                ***************************************:********************

TAIR10          ELLVFPSGCIKKTVNGCRDRTLGKPRSSGLNTDDIHTSSLKISKNDTSNGLTMTTALVEQ 420
992_ref         ELLVFPSGCIKKTVNGCRDRTLGKPRSSGLNTDDIHTSSLKISKNDTSNGLTMTTALVEQ 420
6918_ref        ELLVFPSGCIKKTVNVCRDRTLGKPRSSGLNTDDIHTSSLKISKNDTSNGLTMTTALVEQ 420
6188_altA       ELLVFPSGCIKKTVNVCRDRTLGKPRSSGLNTDDIHTSSLKISKNGTSNGLTMTTALVEQ 420
1061_altA       ELLVFPSGCIKKTVNVCRDRTLGKPRSSGLNTDDIHTSSLKISKNGTSNGLTMTTALVEQ 420
6113_altB       ELLVFPSGCIKKTVNGCRDRTLGKPRSSGLNTDDIHTSSLKISKNDTSNGLTMTTALVEQ 420
6191_altB       ELLVFPSGCIKKTVNGCRDRTLGKPRSSGLNTDDIHTSSLKISKNDTSNGLTMTTALVEQ 420
                *************** *****************************.**************

TAIR10          DAMESLLQGKTSACGAADKGKTREMHVNSTVIYLSDSDEPSSIEYLNGDNLTQVESGSAL 480
992_ref         DAMESLLQGKTSACGAADKGKTREMHVNSTVIYLSDSDEPSSIEYLNGDNLTQVESGSAL 480
6918_ref        DAMESLLQGKTSACGAADKGKTREMHVNSTVIYLSDSDEPSSIEYLNGDNLTQVESGSAL 480
6188_altA       DAMESLLQGKTSACGAADKGKTREMHVNSTVIYLSDSDEPSSIEYLNGDNLTQVKSGSAL 480
1061_altA       DAMESLLQGKTSACGAADKGKTREMHVNSTVIYLSDSDEPSSIEYLNGDNLTQVKSGSAL 480
6113_altB       DAMESLLQGKTSACGAADKGKTREMHVNSTVIYLSDSDEPSSIEYLNGDNLTQVESGSAL 480
6191_altB       DAMESLLQGKTSACGAADKGKTREMHVNSTVIYLSDSDEPSSIEYLNGDNLTQVESGSAL 480
                ******************************************************:*****
```

```
TAIR10       SSGGNEGIVSLDLNNPTKSTKRKGKRVTRTAVQEQNKRSICFFIGEPLSCEEAQERWRWR 540
992_ref      SSGGNEGIVSLDLNNPTKSTNRKGKRVTRTAVQEQNKRSICFFIGEPLSCEEAQERWRWR 540
6918_ref     SSGGNEGIVSLDLNNPTKSTKRKGKRVTRTAVQEQNKRSICFFIGEPLSCEEAQERWRWR 540
6188_altA    SSGGNEGIVSLDLNNPTKSTKRKGKRVTRTAVQEQNKRSICFFIGEPLSCEEAQERWRWR 540
1061_altA    SSGGNEGIVSLDLNNPTKSTKRKGKRVTRTAVQEQNKRSICFFIGEPLSCEEAQERWRWR 540
6113_altB    SSGGNEGIVSLDLNNPTKSTKRKGKRVTRTAVQEQNKRSICFFIGEPLSCEEAQERWRWR 540
6191_altB    SSGGNEGIVSLDLNNPTKSTKRKGKRVTRTAVQEQNKRSICFFIGEPLSCEEAQERWRWR 540
             ********************:***************************************

TAIR10       YELKERKSKSRGQQSEDDEDKIVANVECHYSQAKVDGHTFSLGDFAYIKGEEEETHVGQI 600
992_ref      YELKERKSKSRGQQSEDDEDKIVANVECHYSQAKVDGHTFSLGDFAYIKGEEEETHVGQI 600
6918_ref     YELKERKSKSRGQQSEDDEDKIVANVECHYSQAKVDGHTFSLGDFAYIKGEEEETHVGQI 600
6188_altA    YELKERKSKSRGQQSEDDEDKIVANVECHYSQAKVDGHTFSLGDFAYIKGEEEETHVGQI 600
1061_altA    YELKERKSKSRGQQSEDDEDKIVANVECHYSQAKVDGHTFSLGDFAYIKGEEEETHVGQI 600
6113_altB    YELKERKSKSRGQQSEDDEDKIVANVECHYSQAKVDGHTFSLGDFACIKGEEEETHVGQI 600
6191_altB    YELKERKSKSRGQQSEDDEDKIVANVECHYSQAKVDGHTFSLGDFACIKGEEEETHVGQI 600
             ********************************************** *************
                                   BAH Domain              Y586C
TAIR10       VEFFKTTDGESYFRVQWFYRATDTIMERQATNHDKRRLFYSTVMNDNPVDCLISKVTVLQ 660
992_ref      VEFFKTTDGESYFRVQWFYRATDTIMERQATNHDKRRLFYSTVMNDNPVDCLISKVTVLQ 660
6918_ref     VEFFKTTDGESYFRVQWFYRATDTIMERQATNHDKRRLFYSTVMNDNPVDCLISKVTVLQ 660
6188_altA    VEFFKTTDGESYFRVQWFYRATDTIMERQATNHDKRRLFYSTVMNDNPVDCLISKVTVLQ 660
1061_altA    VEFFKTTDGESYFRVQWFYRATDTIMERQATNHDKRRLFYSTVMNDNPVDCLISKVTVLQ 660
6113_altB    VEFFKTTDGESYFRVQWFYRATDTIMERQATNHDKRRLFYSTVMNDNPVDCLISKVTVLQ 660
6191_altB    VEFFKTTDGESYFRVQWFYRATDTIMERQATNHDKRRLFYSTVMNDNPVDCLISKVTVLQ 660
             ************************************************************

TAIR10       VSPRVGLKPNSIKSDYYFDMEYCVEYSTFQTLRNPKTSENKLECCADVVPTESTESILKK 720
992_ref      VSPRAGLKPNSIKSDYYFDMEYCVEYSTFQTLRNPKTSENKLECWADVVPTESTESILKK 720
6918_ref     VSPRAGLKPNSIKSDYYFDMEYCVEYSTFQTLRNPKTSENKLECCADVVPTKSTESILKK 720
6188_altA    VSPRAGLKPNSIKSDYYFDMEYCVEYSTFQTLRNPKTSENKLECCADVVPTESTESILKK 720
1061_altA    VSPRAGLKPNSIKSDYYFDMEYCVEYSTFQTLRNPKTSENKLECCADVVPTESTESILKK 720
6113_altB    VSPRAGLKPNSIKSDYYFDMEYCVEYSTFQTLRNPKTSENKLECCADVVPTKSTESILKK 720
6191_altB    VSPRAGLKPNSIKSDYYFDMEYCVEYSTFQTLRNPKTSENKLECCADVVPTKSTESILKK 720
             ****.*************************************** ******:********

TAIR10       KSFSGELPVLDLYSGCGGMSTGLSLGAKISGVDVVTKWAVDQNTAACKSLKLNHPNTQVR 780
992_ref      KSFSGELPVLDLYSGCGGMSTGLSLGAKISGVDVVTKWAVDQNTAACKSLKLNHPNTQVR 780
6918_ref     KSFSGELPVLDLYSGCGGMSTGLSLGAKISGVDVVTKWAVDQNTAACKSLKLNHPNTQVR 780
6188_altA    KSFSGELPVLDLYSGCGGMSTGLSLGAKISGVDVVTKWAVDQNKAACKSLKLNHPNTQVR 780
1061_altA    KSFSGELPVLDLYSGCGGMSTGLSLGAKISGVDVVTKWAVDQNKAACKSLKLNHPNTQVR 780
6113_altB    KSFSGELPVLDLYSGCGGMSTGLSLGAKISGVDVVTKWAVDQNTAACKSLKLNHPNTQVR 780
6191_altB    KSFSGELPVLDLYSGCGGMSTGLSLGAKISGVDVVTKWAVDQNTAACKSLKLNHPNTQVR 780
             *******************************************.****************

TAIR10       NDAAGDFLQLLKEWDKLCKRYVFNNDQRTDTLRSVNSTKETSGSSSSSDDDSDSEEYEVE 840
992_ref      NDAAGDFLQLLKEWDKLCKRYVFNNDQRTDTLRSVNSTKETSESSSSSDDDSDSEEYEVE 840
6918_ref     NDAAGDFLQLLKEWDKLCKRYVFNNDQRTDTLRSVNSTKETSESSSSSDDDSDSEEYEVE 840
6188_altA    NDAAGDFLQLLKEWDKLCKRYVFNNDQRTDTLRSVNSTKETSESSSSSDDDSDSEEYEVE 840
1061_altA    NDAAGDFLQLLKEWDKLCKRYVFNNDQRTDTLRSVNSTKETSESSSSSDDDSDSEEYEVE 840
6113_altB    NDSAGDFLQLLKEWDKLCKRYVFNNDQRTDTLRSVNSTKETSESSSSSDDDSDSEEYEVE 840
6191_altB    NDSAGDFLQLLKEWDKLCKRYVFNNDQRTDTLRSVNSTKETSESSSSSDDDSDSEEYEVE 840
             **:*************************************** *****************
                                  CHROMO Domain
TAIR10       KLVDICFGDHDKTGKNGLKFKVHWKGYRSDEDTWELAEELSNCQDAIREFVTSGFKSKIL 900
992_ref      KLVDICFGDPDKTGKNGLKFKVHWKGYRSDEDTWELAEELSNCQDAIREFVTSGFKSKIL 900
6918_ref     KLVDICFGDPDKTGKNGLKFKVHWKGYRSDEDTWELAEELSNCQDAIREFVTSGFKSKIL 900
6188_altA    KLVDICFGDPDKTGKNGLKFKVHWKGYRSDEDTWELAEELSNCQDAIREFVTSGFKSKIL 900
1061_altA    KLVDICFGDPDKTGKNGLKFKVHWKGYRSDEDTWELAEELSNCQDAIREFVTSGFKSKIL 900
6113_altB    KLVDICFGDPDKTGKNGLKFKVHWKGYRSDEDTWELAEELSNCQDAIREFVTSGFKSKIL 900
6191_altB    KLVDICFGDPDKTGKNGLKFKVHWKGYRSDEDTWELAEELSNCQDAIREFVTSGFKSKIL 900
             ********* **************************************************
                         C5 Methyltransferase Domain
TAIR10       PLPGRVGVICGGPPCQGISGYNRHRNVDSPLNDERNQQIIVFMDIVEYLKPSYVLMENVV 960
992_ref      PLPGRVGVICGGPPCQGISGYNRHRNVDSPLNDERNQQIIVFMDIVEYLKPSYVLMENVV 960
6918_ref     PLPGRVGVICGGPPCQGISGYNRHRNVDSPLNDERNQQIIVFMDIVEYLKPSYVLMENVV 960
6188_altA    PLPGRVGVICGGPPCQGISGYNRHRNVDSPLNDERNQQIIVFMDIVEYLKPSYVLMENVV 960
1061_altA    PLPGRVGVICGGPPCQGISGYNRHRNVDSPLNDERNQQIIVFMDIVEYLKPSYVLMENVV 960
6113_altB    PLPGRVGVICGGPPCQGISGYNRHRNVDSPLNDEKNQQIIVFMDIVEYLKPSYVLMENVV 960
6191_altB    PLPGRVGVICGGPPCQGISGYNRHRNVDSPLNDEKNQQIIVFMDIVEYLKPSYVLMENVV 960
             **********************************:*************************
                                               R934K
TAIR10       DILRMDKGSLGRYALSRLVNMRYQARLGIMTAGCYGLSQFRSRVFMWGAVPNKNLPPFPL 1020
992_ref      DILRMDKGSLGRYALSRLVNMRYQARLGIMTAGCYGLSQFRSRVFMWGAVPNKNLPPFPL 1020
6918_ref     DILRMDKGSLGRYALSRLVNMRYQARLGIMTAGCYGLSQFRSRVFMWGAVPNKNLPPFPL 1020
6188_altA    DILRMDKGSLGRYALSRLVNMRYQARLGIMTAGCYGLSQFRSRVFMWGAVPNKNLPPFPL 1020
1061_altA    DILRMDKGSLGRYALSRLVNMRYQARLGIMTAGCYGLSQFRSRVFMWGAVPNKNLPPFPL 1020
6113_altB    DILRMDKGSLGRYALSRLVNMRYQARLGIMTAGCYGLSQFRSRVFMWGAVPNKNLPPFPL 1020
6191_altB    DILRMDKGSLGRYALSRLVNMRYQARLGIMTAGCYGLSQFRSRVFMWGAVPNKNLPPFPL 1020
             ************************************************************
```

```
TAIR10          PTHDVIVRYGLPLEFERNVVAYAEGQPRKLEKALVLKDAISDLPHVSNDEDREKLPYESL 1080
992_ref         PTHDVIVRYGLPLEFERNVVAYAEGQPRKLEKALVLKDAISDLPHVSNDEDREKLPYESL 1080
6918_ref        PTHDVIVRYGLPLEFERNVVAYAEGQPRKLEKALVLKDAISDLPHVSNDEDREKLPYESL 1080
6188_altA       PTHDVIVRYGLPLEFERNVVAYAEGQPRKLEKALVLKDAISDLPHVSNDEDREKLPYESL 1080
1061_altA       PTHDVIVRYGLPLEFERNVVAYAEGQPRKLEKALVLKDAISDLPHVSNDEDREKLPYESL 1080
6113_altB       PTHDVIVRYGLPLEFERNVVAYAEGQPRKLEKALVLKDAISDLPHVSNEEDREKLPYESL 1080
6191_altB       PTHDVIVRYGLPLEFERNVVAYAEGQPRKLEKALVLKDAISDLPHVSNEEDREKLPYESL 1080
                ************************************************:***********
                                                             D1068E
TAIR10          PKTDFQRYIRSTKRDLTGSAIDNCNKRTMLLHDHRPFHINEDDYARVCQIPKRKGANFRD 1140
992_ref         PKTDFQRYIRSTKRDLTGSAIDNCNKRTMLLHDHRPFHINEDDYARVCQIPKRKGANFRD 1140
6918_ref        PKTDFQRYIRSTKRDLTGSAIDNCNKRTMLLHDHRPFHINEDDYARVCQIPKRKGANFRD 1140
6188_altA       PKTDFQRYIRSTKRDLTGSAIDNCNKRTMLLHDHRPFHINEDDYARVRQIPKRKGANFRD 1140
1061_altA       PKTDFQRYIRSTKRDLTGSAIDNCNKRTMLLHDHRPFHINEDDYARVRQIPKRKGANFRD 1140
6113_altB       PKTDFQRYIRSTKRDLTGSAIDNCNKRTMLLHDHRPFHINEDDYARVCQIPKRKGANFRD 1140
6191_altB       PKTDFQRYIRSTKRDLTGSAIDNCNKRTMLLHDHRPFHINEDDYARVCQIPKRKGANFRD 1140
                *********************************************** ************
                                                             C1127R
TAIR10          LPGLIVRNNTVCRDPSMEPVILPSGKPLVPGYVFTFQQGKSKRPFARLWWDETVPTVLTV 1200
992_ref         LPGIIVRNNTVCRDPSMEPVILPSGKPLVPGYVFTFQQGKSKRPFARLWWDETVPTVLTV 1200
6918_ref        LPGLIVRNNTVCRDPSMEPVILPSGKPLVPGYVFTFQQGKSKRPFARLWWDETVPTVLTV 1200
6188_altA       LPGLIVRNNTVCRDPSMEPVILPSGKPLVPGYVFTFQQGKSKRPFARLWWDETVPTVLTV 1200
1061_altA       LPGLIVRNNTVCRDPSMEPVILPSGKPLVPGYVFTFQQGKSKRPFARLWWDETVPTVLTV 1200
6113_altB       LPGLIVRNNTVCRDPSMEPVILPSGKPLVPGYVFTFQQGKSKRPFARLWWDETVPTVLTV 1200
6191_altB       LPGLIVRNNTVCRDPSMEPVILPSGKPLVPGYVFTFQQGKSKRPFARLWWDETVPTVLTV 1200
                ***:********************************************************

TAIR10          PTCHSQALLHPEQDRVLTIRESARLQGFPDYFQFCGTIKERYCQIGNAVAVSVSRALGYS 1260
992_ref         PTCHSQALLHPEQDRVLTIRESARLQGFPDYFQFCGTIKERYCQIGNAVAVSVSRALGYS 1260
6918_ref        PTCHSQALLHPEQDRVLTIRESARLQGFPDYFQFCGTIKERYCQIGNAVAVSVSRALGYS 1260
6188_altA       PTCHSQALLHPEQDRVLTIRESARLQGFPDYFQFCGTIKERYCQIGNAVAVSVSRALGYS 1260
1061_altA       PTCHSQALLHPEQDRVLTIRESARLQGFPDYFQFCGTIKERYCQIGNAVAVSVSRALGYS 1260
6113_altB       PTCHSQALLHPEQDRVLTIRESARLQGFPDYFQFCGTIKERYCQIGNAVAVSVSRALGYS 1260
6191_altB       PTCHSQALLHPEQDRVLTIRESARLQGFPDYFQFCGTIKERYCQIGNAVAVSVSRALGYS 1260
                ************************************************************

TAIR10          LGMAFRGLARDEHLIKLPQNFSHSTYPQLQETIPH 1295
992_ref         LGMAFRGLARDEHLIKLPQNFSHSTYPQLQETIPH 1295
6918_ref        LGMAFRGLARDEHLIKLPQNFSHSTYPQLQETIPH 1295
6188_altA       LGMAFRGLARDEHLIKLPQNFSHSTYPQLQETIPH 1295
1061_altA       LGMAFRGLARDEHLIKLPQNFSHSTYPQLQETIPH 1295
6113_altB       LGMAFRGLARHEHLIKLPQNFSHSTYPQLQEAIPH 1295
6191_altB       LGMAFRGLARHEHLIKLPQNFSHSTYPQLQEAIPH 1295
                **********.********************:***
```